\documentclass{article}

\usepackage[preprint,nonatbib]{neurips_2026}

\usepackage[utf8]{inputenc} 
\usepackage{url}            

\usepackage{amsfonts}       
\usepackage{nicefrac}       

\usepackage{microtype}      
\usepackage{xcolor}         
\usepackage{float}
\usepackage{graphicx}    
\usepackage{array}       
\usepackage{booktabs,multirow,makecell}
\usepackage{subcaption}
\usepackage{wrapfig}

\usepackage[normalem]{ulem}

\usepackage[all,british]{foreign}

\renewcommand\foreignabbrfont{\normalfont}
\DeclareRobustCommand{\aka}{\xperiodafter{{\foreignabbrfont{a.k.a}}}}
\usepackage{enumitem}
\usepackage[xindy]{glossaries-extra}
\setabbreviationstyle[acronym]{long-short}
\newacronym{DFT}{DFT}{Density functional theory}
\usepackage{xspace}
\DeclareRobustCommand{\datasetname}{\textsc{CataLiUst Ti2C-MXene}\xspace}
\DeclareRobustCommand{\fivetimesish}{$\approx\!5\times$\xspace}
\usepackage{hyperref}       
\usepackage[capitalise,noabbrev]{cleveref}
\crefname{equation}{Eq.}{Eqs.}
\Crefname{equation}{Equation}{Equations}

\usepackage[numbers]{natbib}

\title{\datasetname Dataset}
\title{Benchmark Dataset for Catalysis on 2D MXenes}

\author{Pavlo Melnyk$^{\star 1}$ \quad Anmar Karmush$^{\star 1}$ \quad \textbf{Mårten Wadenbäck}$^{1}$ \\ \vspace{2pt}
\textbf{Ania Beatriz Rodríguez-Barrera}$^{2,3}$ ~ \textbf{Johanna Rosen}$^{2,3}$ ~ \textbf{Michael Felsberg}$^{1}$ ~ \textbf{Jonas Björk}$^{\star \dagger 2,3}$\\
{\small $^1$Computer Vision and Learning Systems, Department of Electrical Engineering (ISY)~\&~AI4X,}\\
{\small$^2$Materials Design Division, Department of Physics, Chemistry and Biology (IFM), and} \\
{\small $^3$Wallenberg Initiative Materials Science for Sustainability (WISE);} \\
{\small Linköping University (LiU), 58183 Linköping, Sweden.}\\
{\small$\star$ equal contribution ~~ $\dagger$ corresponding author}~
{\small \texttt{jonas.bjork@liu.se}}
}

\begin{document}
\maketitle
\vspace{-20pt}
\begin{abstract}\vspace{-5pt}
  Merging first-principles calculations with machine learning (ML), we aim to accelerate the exploration of catalytic behaviour in novel materials. We focus on two-dimensional (2D) Ti$_2$CT$_y$ MXenes, whose versatile surface chemistry makes them particularly compelling candidates for catalysis. 
Resolving their composition and structure under realistic conditions exceeds the reach of standard density functional theory (DFT) due to computational cost.
To address this challenge, we generate a comprehensive dataset of 50,000 DFT calculations for training and 10,000 for testing, encompassing both Ti$_2$CT$_y$ MXene configurations and molecular systems, along with an additional test dataset with 1000 genuinely new, larger systems to investigate how well models generalise.
We train and validate widely used and competitive machine learning interatomic potentials (MLIP) models, EquiformerV2, MACE, MatRIS, UPET, and MatRIS that accurately predict atomic forces and formation energies --- quantities that DFT must repeatedly compute for structural and catalytic investigations --- for these 2D materials. 
This combined DFT–ML framework achieves computational acceleration of the order ${\sim}1-4\cdot10^3$ (on a CPU) while maintaining desired-level accuracy (${\sim} {\pm} 10$ meV/Å for forces and ${\sim} {\pm} 1$ meV for per-atom energies), paving the way for more efficient investigations of MXene catalytic behaviour.
Moreover, we perform an extensive qualitative evaluation of the trained models, showcasing the importance of the comprehensive simulation-based comparison beyond the benchmark metrics.
The dataset and the trained models with the code are available at \url{https://huggingface.co/datasets/CatalystAnonymous/catalyst_mxenes}.
\end{abstract}
\vspace{-10pt}
\begin{wrapfigure}{R}{0.5\textwidth}
\includegraphics[width=0.5\textwidth]{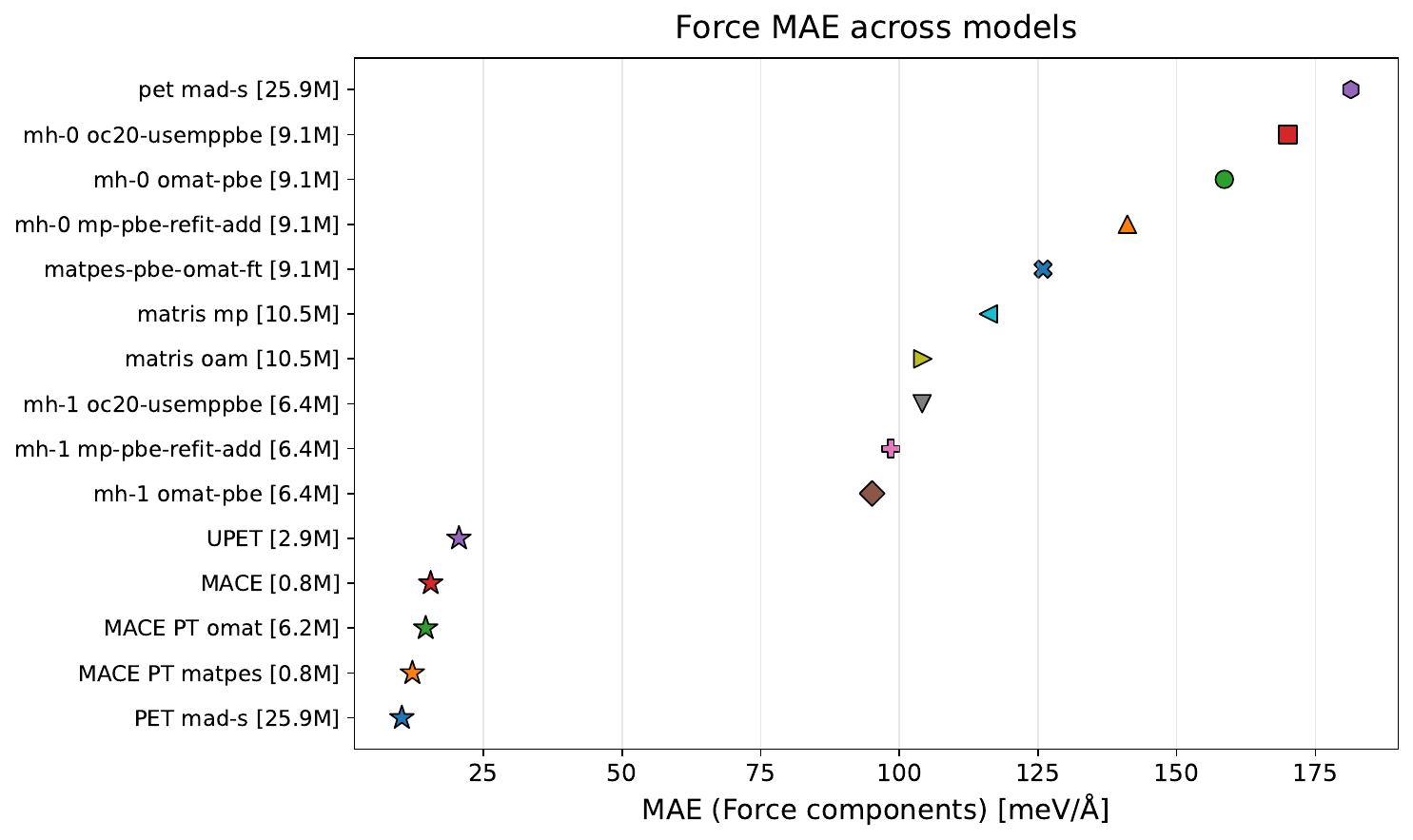}
\caption{Mean absolute error (MAE) of force predictions across publicly available foundation models given Ti$_2$CT$_y$ MXene systems.
The results indicate limited transferability of existing foundation models to MXene catalysis data, hence the need for the proposed \datasetname dataset. 
Star markers denote models trained from scratch and foundation models fine-tuned on the proposed dataset.
}
\label{fig:mace_intro}
\end{wrapfigure}
\section{Introduction}
\vspace{-5pt}
A central challenge in computational catalysis is balancing accuracy and efficiency. Density functional theory (DFT) provides a reliable description of surface chemistry, delivering energies and atomic forces that can be used to model molecule-surface interactions and reaction pathways. However, the computational cost of obtaining such data restricts simulations to relatively small systems and short timescales --- far from resembling realistic operating conditions of catalysts. This limitation is particularly problematic because catalysts are inherently dynamic: their structure and reactivity evolve under reaction conditions, and idealised models may risk missing the relevant chemistry.

Machine learning (ML) interatomic potentials (MLIP) \cite{wang2024machine}, often within the framework of geometric deep learning (GDL) \cite{bronstein2017geometric, bronstein2021geometric}, offer a promising way forward by approximating DFT-calculated energies and forces with near-DFT accuracy while operating orders of magnitude faster \cite{MLIP_DFT_1, MLIP_DFT_2, mattersim, batzner20223}. This facilitates the study of catalytic systems under more realistic conditions, helping to bridge the gap between idealised DFT models and experimental environments. Such simulations can, in turn, identify structural motifs and environments that are truly representative, guiding more focused and accurate mechanistic studies at the DFT level.

While the existing benchmarks for related catalysis problems \cite{Chanussot2021, Tran2023} involve metal oxides and metal-intermetallics, another group of materials for which the above considerations are particularly relevant is that of two-dimensional (2D) transition metal carbides (MXenes). 
The surfaces of these materials are far from well-defined, as sluggish adsorption processes during synthesis lead to a variety of stable surface functionalisations with distinct chemical reactivities. 
At the same time, experimental studies have demonstrated promising catalytic performance for MXenes, and DFT investigations have shown that the catalytic activity is highly sensitive to the precise surface chemical environment. For example, for MXenes with mixed O and OH terminations, the dehydrogenation of alkanes to olefins has a reactivity that decreases linearly with increasing number of OH terminations \cite{Niu2020, Niu2021}. 
At the same time, MXenes with high OH coverage have been predicted to have a high reactivity to CO$_2$ reduction \cite{Parui2022}. 
Thus, even for MXenes in their idealised, well-defined form, the catalytic activity can be systematically tuned by modifying the surface terminations. 

However, despite their great promise in catalysis, very little is known about the actual state of MXenes under operating conditions. 
Capturing how their surface terminations and reactivity evolve in reactive environments requires simulations that are both accurate and computationally feasible --- beyond the scope of conventional DFT approaches. Although there have been ML approaches applied to catalysis on MXenes --- for example, to predict MXene-supported single-atom catalysts for oxygen reduction and evolution reactions~\cite{Guo2025} as well as ammonia synthesis~\cite{Lin2025} --- most work has focused on static screening of catalytic activity. 

The application of ML models to capture the dynamic surface chemistry and evolving reactivity of MXenes under operating conditions has largely been overlooked. 
Moreover, existing MLIP foundation models, such as MACE~\cite{batatia2022mace, batatia2023foundation}, PET~\cite{PET-MAD-1.5-2026}, and MatRIS~\cite{zhou2026matris}, while highly successful across a wide range of bulk and surface materials, exhibit limited transferability to MXenes, as can be seen in \cref{fig:mace_intro}. 
This can be attributed to the complex and heterogeneous surface terminations of MXenes, which are underrepresented in the datasets used to train current foundation models. 
As a result, these models often fail to provide reliable energies and forces for MXene systems, particularly when surface chemistry is involved. 
In this context, equivariant (\ie symmetry-aware) MLIPs trained on high-quality DFT data provide a powerful route to explore the dynamic surface chemistry of MXenes and to establish a more realistic understanding of their catalytic behaviour under working conditions.

We summarise our contributions as follows:

\noindent\textbf{(i)} 
We introduce \datasetname, a dataset for catalysis on Ti$_2$CT$_y$ MXenes, comprising 50,000 DFT calculations for training and 10,000 for testing. 
The dataset includes MXene configurations with and without molecular adsorbates, and features an additional test set with 1,000 \fivetimesish larger atomic systems to explicitly probe model generalisation to the input size.

\noindent\textbf{(ii)} 
We train widely used and competitive interatomic potential models, EquiformerV2 \cite{Equiformerv2}, MACE~\cite{batatia2022mace}, (U)PET~\cite{UPET-2026}, and MatRIS~\cite{zhou2026matris}, on the proposed dataset, and systematically evaluate their out‑of‑distribution generalisation, including to larger atomic systems.

\noindent\textbf{(iii)} 
We integrate the trained models into a computational catalysis workflow, achieving 1,270--4,258-fold acceleration (on a CPU) over the DFT method while maintaining desired accuracy and robustness, thereby enabling efficient large‑scale studies of MXene catalytic behaviour.
\begin{figure}
\centering
\includegraphics[width=0.9\linewidth]{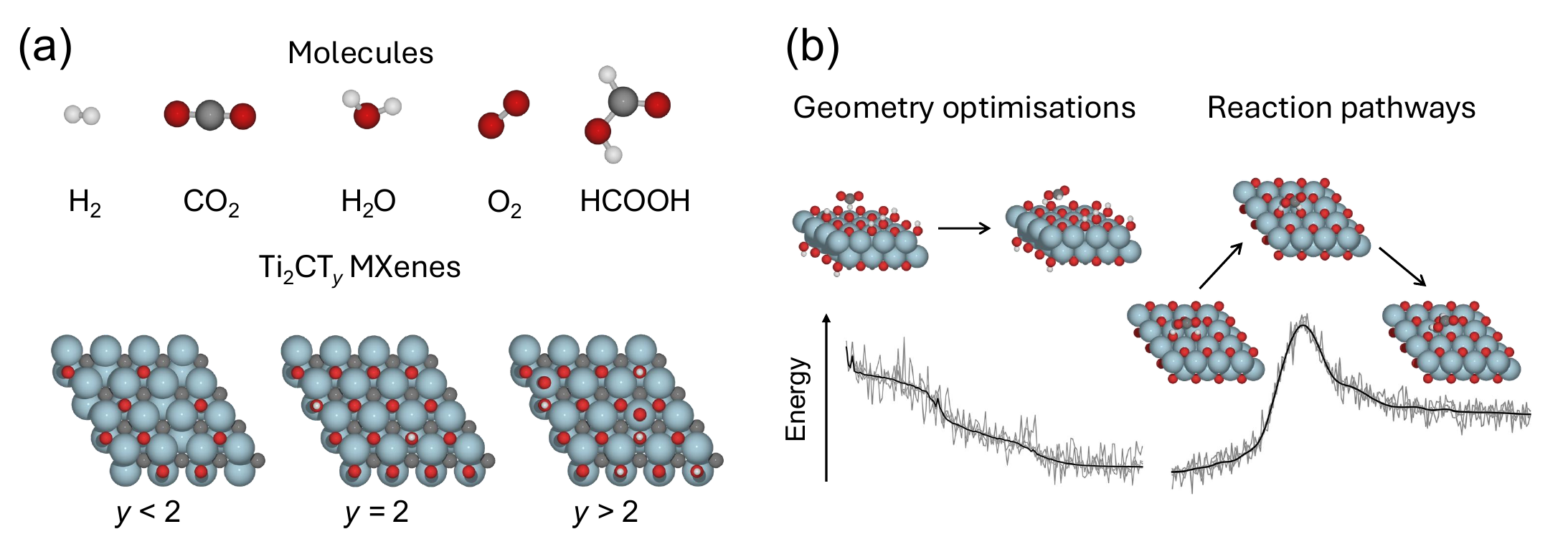}
\caption{Types of (a) systems and (b) DFT calculations included in the dataset.
It comprises DFT calculations of five different molecules adsorbed on 
Ti$_2$CT$_y$ MXenes with varying surface termination configurations 
(denoted T$_y$). The calculations include, for instance, trajectories from 
geometry optimisations and reaction pathways, as well as rattled geometries
along these trajectories, providing a diverse set of configurations for 
training. Atomic simulation environment (ASE)~\cite{ase-paper} was used 
for structural visualisation.
}\vspace{-15pt}
\label{fig:training data}
\end{figure}
\vspace{-15pt}
\section{Background}
\vspace{-5pt}
\subsection{Density Functional Theory (DFT)}
\vspace{-5pt}
DFT is a first-principles method: given only the atomic species and coordinates, it computes the total energy, forces on atoms, and other properties derived from the electron density. At its core, DFT replaces the many-electron wave function with the electron density—a simpler quantity that, in principle, contains all ground-state information about the system \cite{Hohenberg1964, Kohn1965}. 

In practice, DFT is approximate at several levels. 
The primary challenge lies in describing the non-classical part of the electron-electron interaction, namely, the exchange and correlation effects. 
These contributions are encompassed in the so-called exchange–correlation (XC) functional. 
Since the exact form of this functional is unknown, a range of approximations exist, each balancing accuracy and computational cost. 
The predictive power of DFT therefore depends critically on how well the chosen XC functional represents the relevant electronic interactions.

Accurately describing non-local dispersion forces, or van der Waals (vdW) interactions, is a particular challenge, as generalised gradient approximation (GGA) functionals cannot capture them. 
To address this, we employ the vdW density functional (vdW-DF)~\cite{Dion2004} in the form introduced by \cite{Hamada2014} (rev-vdW-DF2). This functional accurately describes structural parameters of MXenes~\cite{Niu2025}, molecular adsorption energies and distances~\cite{Chen2025}, lattice parameters and interlayer binding energies in graphite~\cite{Hamada2014}, and other weakly bonded layered systems~\cite{Tran2019}.

When comparing energies across different DFT datasets, the choice of zero-energy reference is crucial; this is discussed in Appendix~\ref{app:dft_notes}.
\vspace{-7pt}
\subsection{Equivariant Modelling with Geometric Deep Learning}\vspace{-7pt}
Incorporating correct inductive biases in learning systems facilitates the learning process and enables tractable learning as the space of transformations acting on the input grows \cite{bronstein2021geometric}.
In general, \textit{symmetries}, both as (i) the transformations of an object (or its properties) that leave it unchanged (\textit{invariance}) and (ii) the transformations that change it in a predictable manner (\textit{equivariance}), are powerful inductive biases \cite{van1995vision, weiler2023equivariant}.
Note that (i) is a special case of (ii).
More formally, GDL \cite{bronstein2021geometric} uses the language of group and representation theory and provides a framework for constructing neural network (NN) architectures that adhere to the symmetries in the data the NNs are to model. 

For instance, for a classification model taking in an input 3D point cloud representing the surface of an object, rotating the point cloud should not affect the class assignment produced by the model (invariance under rotations \aka SO($3$)-invariance) \cite{esteves17_learn_so_equiv_repres_with_spher_cnns,weiler20183d, anderson2019cormorant, fuchs2020se3, tetrasphere}. 
At the same time, a model predicting forces acting on the object should output correspondingly rotated force vectors if the input point cloud is rotated (SO($3$)-equivariance).
When working with 3D data, it is often also beneficial to be able to handle translations and reflections (mirroring), which, along with rotations, constitute the Euclidean group E($n$). 
In such cases, we refer to models as E($3$)-equivariant (or -invariant).
In the following section, we show how the GDL framework is used in our work for catalysis on 2D MXenes.
\section{Method}
\vspace{-3pt}
We aim to enable accelerated computational investigations of catalytic processes on 2D MXenes. 
To achieve this, we use first-principles DFT calculations to produce a dataset of MXene systems, which we use to train symmetry-aware models to predict atomic forces and formation energies.
\vspace{-3pt}
\subsection{\datasetname: DFT dataset of molecules interacting with MXene surfaces}
\vspace{-3pt}
While prominent ML catalysis efforts, such as the Open Catalyst Project \cite{Chanussot2021,Tran2023}, focus on screening across a wide range of materials to identify promising candidates, here we adopt a complementary approach.
Instead of targeting broad screening, we focus on a single material --- the Ti$_2$CT$_y$ MXene --- chosen for its catalytic relevance and rich surface chemistry. 
This allows us to construct a highly detailed and physically consistent dataset that captures the complexity of a single catalyst under realistic conditions. 
Our dataset, called \datasetname, spans a large range of surface terminations, from non-terminated to over-terminated configurations, including mixed O and OH coverage.
Furthermore, it includes configurations far from equilibrium obtained from high-temperature molecular dynamics and reaction pathway calculations, enabling the model to capture chemical environments encountered under operating conditions.
The dataset covers molecules relevant to CO$_2$ reduction to formic acid, as well as other key catalytic processes such as the hydrogen evolution reaction (HER) and the oxygen evolution reaction (OER).

The diversity and depth of configurations are made possible through systematic DFT calculations, which form the foundation of the dataset.
Each data point corresponds to a static DFT calculation that provides the total energy of the system and the forces acting on its atoms.
For the energy, we use the \textbf{formation} energy, defined as
\begin{equation}
\Delta_f E = E_\mathrm{DFT}\left (\mathrm{system} \right) - \sum_i{E_\mathrm{ref}\left (\mathrm{atom}_i \right)},
\label{eq:formation-energy}
\end{equation}
where $E_\mathrm{DFT}\left (\mathrm{system} \right)$ is the total DFT energy and 
$E_\mathrm{ref}\left (\mathrm{atom}_i \right)$ is the reference energy of atom $i$ in the 
system, which contains the positions, \ie coordinates, $\mathrm{pos}$ and atomic species 
(numbers) of the atoms. 
Reference energies are taken as the energies of the respective elemental 
phases --- graphitic C, hcp Ti, O$_2$ gas, and H$_2$ gas --- normalised per atom. 

Note that $\Delta_f E$ is the formation energy of the complete unit 
cell representing a system. It can be interpreted as the total 
potential energy of the system with a physically meaningful reference, 
which the raw DFT total energy lacks when using pseudopotential-based 
approaches. 
This allows meaningful comparisons across different structures and chemical environments, while still enabling the 
calculation of energy differences relevant to catalysis, such as 
adsorption, reaction, and activation energies. We discuss this point 
in more detail in the Appendix.
The data, outlined in \cref{fig:training data}, comprise DFT calculations for: 
\begin{itemize}[itemsep=0ex,topsep=0ex]
    \item Individual molecules (H$_2$, CO$_2$, H$_2$O, O$_2$, HCOOH).
\vspace{-2pt}
    \item Ti$_2$CT$_y$ MXenes with various surface termination configurations (denoted T$_y$). 
    Both the coverage of surface terminations (value of $y$) and the types of  terminations (O and/or OH)
    are considered to capture the diversity of MXene surface chemistry.
\vspace{-2pt}
    \item Molecules adsorbed on the different MXene surfaces.
\end{itemize}
The data originate from several calculation types:
\begin{enumerate}[itemsep=0ex,topsep=0ex]
    \item \textbf{Geometry optimisations.}
    Configurations correspond largely to local minima, where atomic forces are generally small.
    \item \textbf{Reaction pathway calculations.}
    These probe reaction mechanisms by optimising along a reaction coordinate (using the nudged elastic band method). Forces are not necessarily small, as these configurations represent specific regions of the potential energy surface connecting local minima.
    \item \textbf{Rattled structures.}
These configurations are generated by applying systematic random displacements (“rattling”) to both local minima, optimisation trajectories, and reaction pathways obtained from NEB calculations. 
By varying the displacement amplitudes, we ensure broad coverage of off-equilibrium configurations with diverse forces. 
\end{enumerate}

Notably, our dataset contains only four atom types, but the complexity of the system is amplified by the diversity of their chemical environments. In contrast to systems with a larger number of distinct elements, where surface and molecular species are inherently distinguished, here each element appears in multiple, chemically distinct bonding motifs. For example, a carbon atom in a MXene --- coordinated to six titanium atoms --- is fundamentally different from a carbon atom in a molecule, and carbon atoms in different molecules also differ significantly in bonding. Likewise, oxygen behaves very differently when bound as a surface termination to titanium atoms compared to when participating in various molecular configurations. This rich variety in the chemical surroundings of different atom types greatly increases the challenge for any model.

To complement the test dataset, we generated larger structures to assess the ability of the models to generalise to extended systems. 
These structures were constructed by stitching together randomly selected $4 \times 4$ surface unit cells from the training set into larger supercells of size $n_1 \times n_2 \times n_3$, where $n_1, n_2 \in \{1,2,3\}$ and $n_3 = 1$. 
To avoid trivial reconstructions, we imposed the constraint that $(n_1, n_2) \neq (1,1)$.
The supercells were assembled from MXene unit cells featuring different surface terminations and adsorbate species. 
As a result, neighbouring tiles exhibit distinct termination patterns, leading to chemically heterogeneous interfaces with abrupt changes in local bonding environments. 
These interfacial regions introduce configurations that are absent from both the training and original test datasets.
Each constructed structure was subsequently subjected to a soft geometry optimisation using density functional theory (DFT). The relaxation was terminated when the maximum force component fell below a threshold of either $0.5~\mathrm{eV/\AA}$ or $1.0~\mathrm{eV/\AA}$, with approximately half of the structures assigned to each criterion. This ensures that the resulting 
 configurations are physically meaningful while remaining out of equilibrium.
 
\paragraph{Details of DFT calculations}
Periodic density functional theory calculations are performed with the VASP code~\cite{Kresse1996}, using the project-augmented wave method~\cite{Blochl1994, Kresse1999}, and with a plane wave basis expanded to a kinetic energy cutoff of 450 eV\@. The MXenes are represented in $p\left(4\times4 \right)$ unit cell together with a $3\times3$ $k$-point sampling, which was adjusted for repeated systems to give a consistent $k$-point density across systems, ensuring numerically converged energies and forces.
\begin{table*}[t]
\centering
\small
\setlength{\tabcolsep}{6pt}
\renewcommand{\arraystretch}{1.2}
\begin{tabular}{l r r r r c}
\toprule
\textbf{Split} &
\textbf{min} &
\textbf{median} &
\textbf{max} &
\textbf{mean $\pm$ std}\phantom{0} &
\textbf{Number of systems} \\
\midrule
Train & 2 & 96.0 & 156 & $\phantom{0}88.4 \pm \phantom{0}27.8$ & 50,000 \\
~~~~Train (used split) & 2 & 98.0 & 156 & $\phantom{0}88.4 \pm \phantom{0}27.7$ & 40,000 \\
~~~~Val (used split) & 2 & 98.0 & 156 & $\phantom{0}88.5 \pm \phantom{0}27.8$ & 10,000 \\
\midrule
Test  & 2 & 96.0 & 158 & $\phantom{0}87.0 \pm \phantom{0}30.4$ & 10,000 \\
\midrule
Test \emph{larger systems} ~(\fivetimesish) & 128 & 338 & 923 & $415.1 \pm 215.1$ & 1,000 \\
\bottomrule
\end{tabular}
\caption{
Statistics of the number of atoms per system in the training and test splits, as well as an additional 1,000 test \emph{larger systems}(\fivetimesish), of the proposed dataset.
}
\label{tab:num_atoms_standard}
\end{table*}
\begin{figure}[t]
    \centering
    \begin{subfigure}[t]{0.4\textwidth}
        \centering
        \includegraphics[width=\linewidth,height=4.6cm,keepaspectratio]{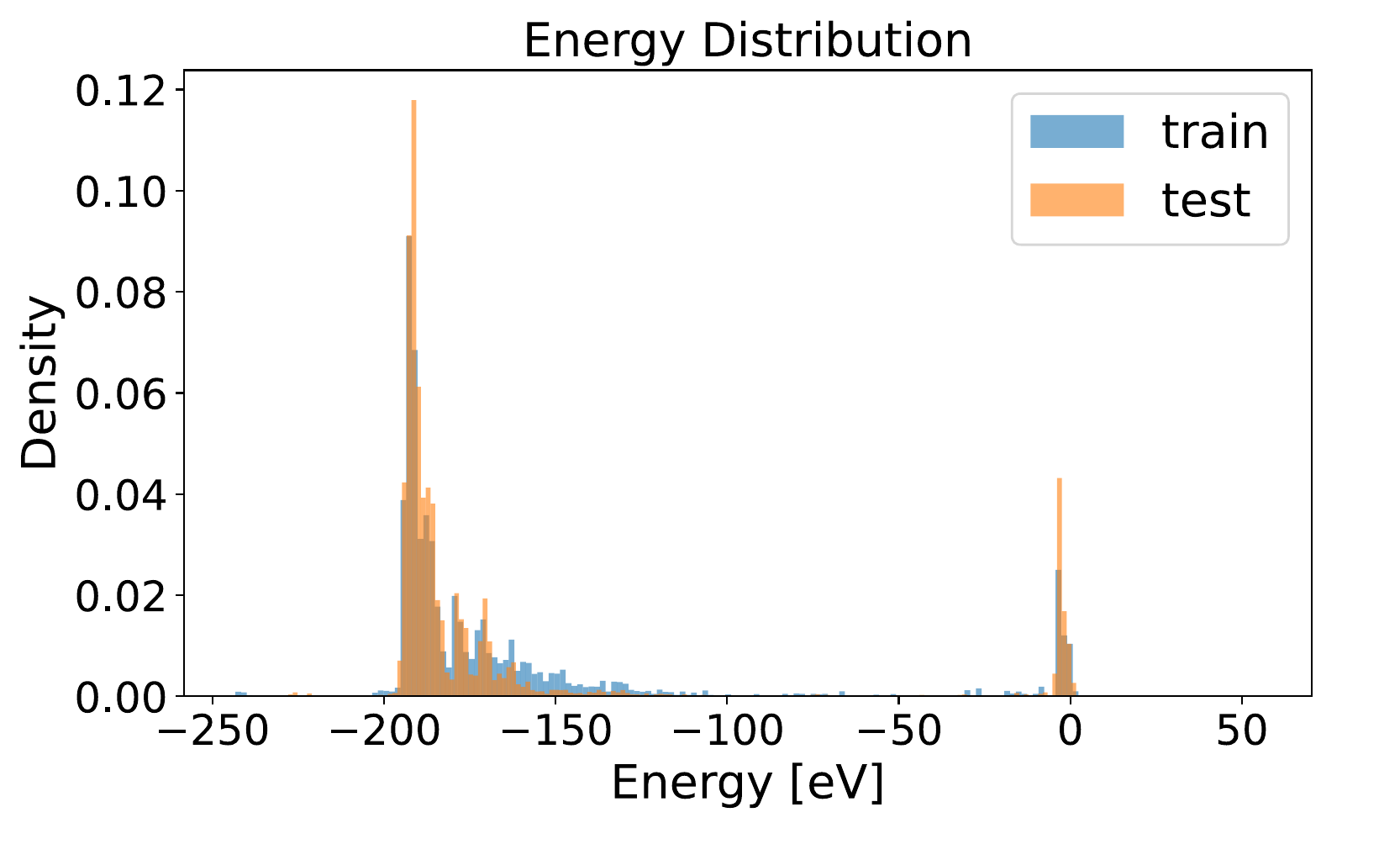}
        \caption{Energies}
        \label{fig:energy_histogram}
    \end{subfigure}\hspace{0.02\textwidth}  
    \begin{subfigure}[t]{0.4\textwidth}
        \centering
        \includegraphics[width=\linewidth,height=4.6cm,keepaspectratio]{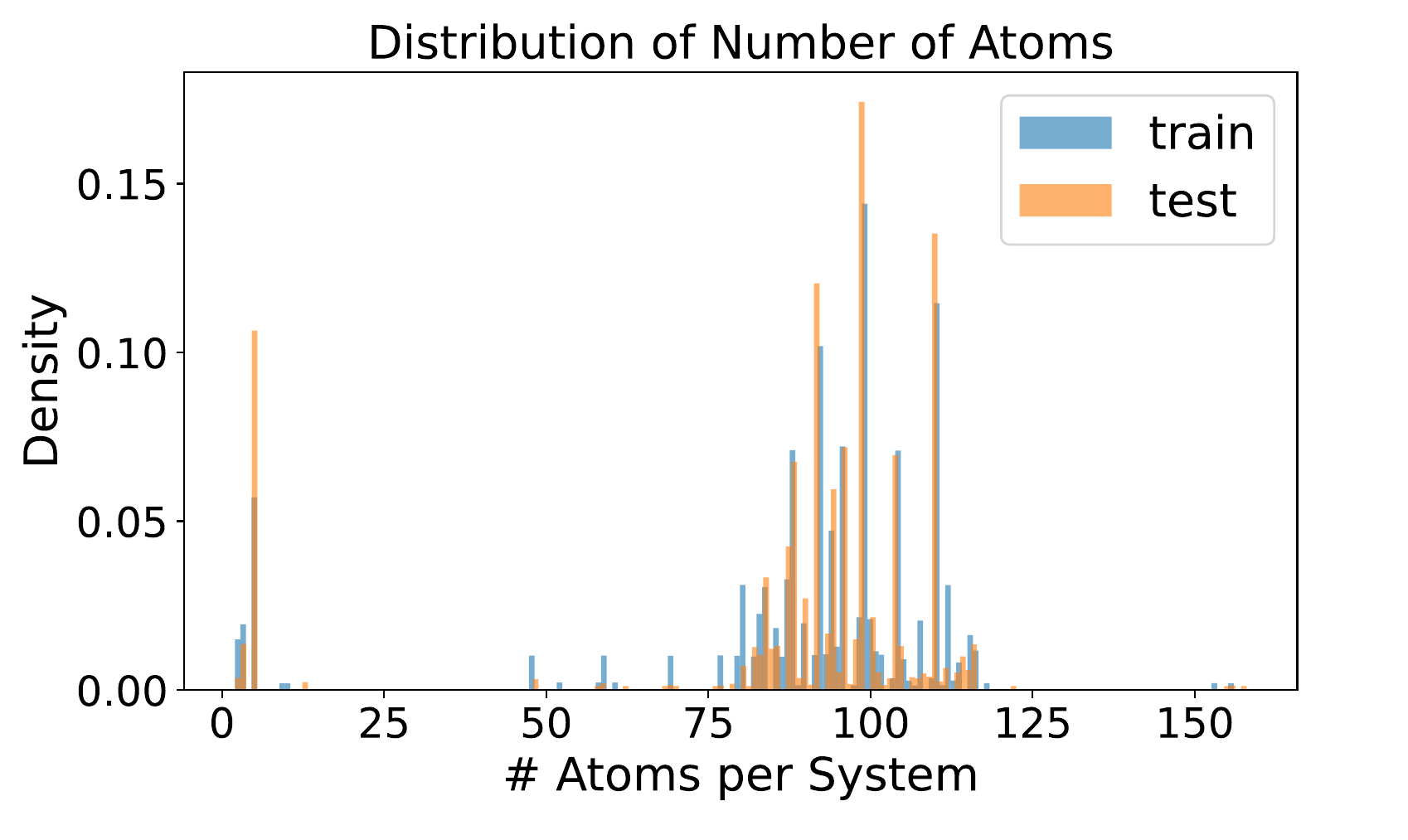}
        \caption{Number of atoms}
        \label{fig:num_atoms_histogram}
    \end{subfigure}

    \medskip 

    \begin{subfigure}[t]{0.4\textwidth}
        \centering
        \includegraphics[width=\linewidth,height=4.6cm,keepaspectratio]{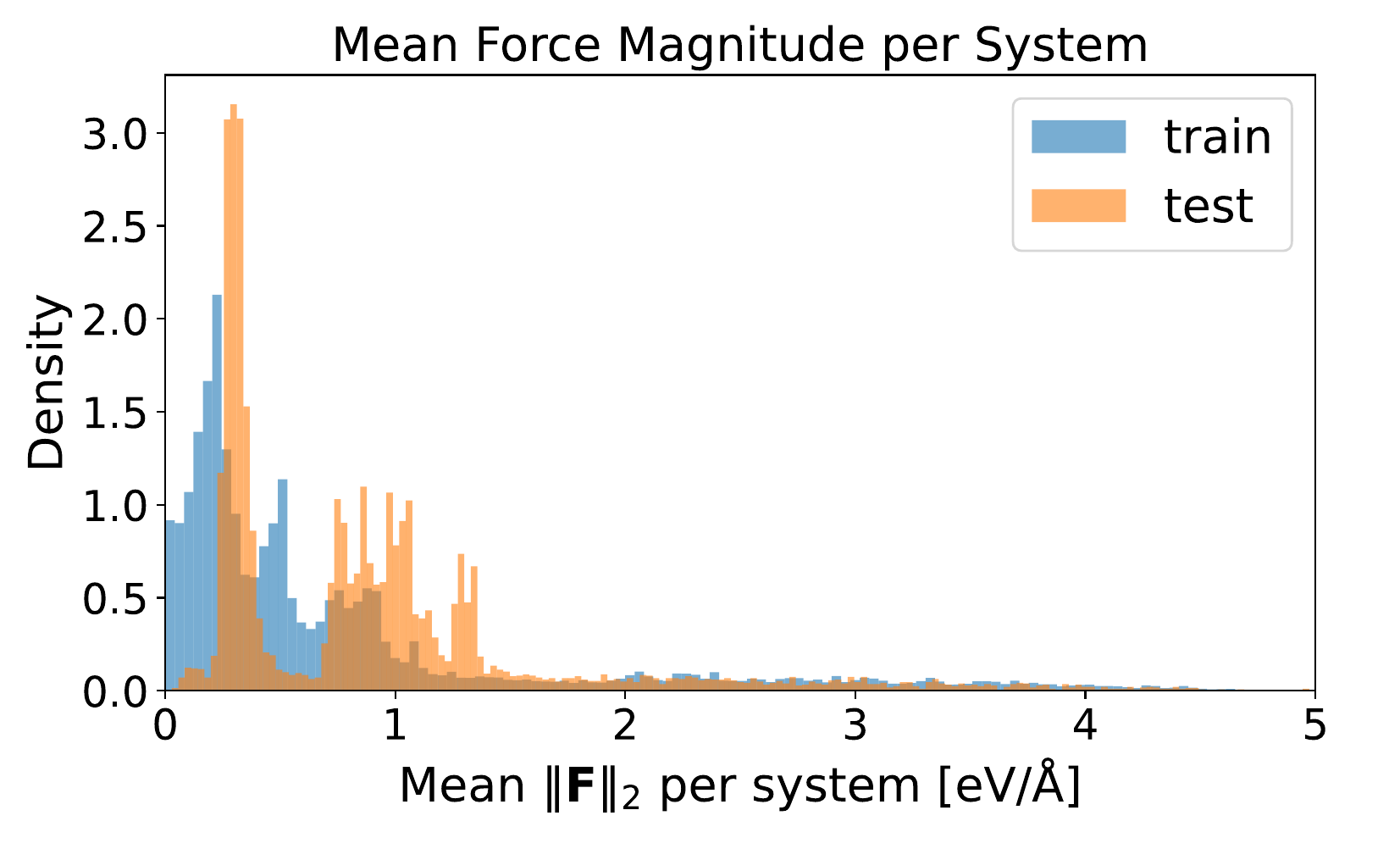}
        \caption{Forces per system}
        \label{fig:force_per_system_histogram}
    \end{subfigure}\hspace{0.02\textwidth}  
    \begin{subfigure}[t]{0.4\textwidth}
        \centering
        \includegraphics[width=\linewidth,height=4.6cm,keepaspectratio]{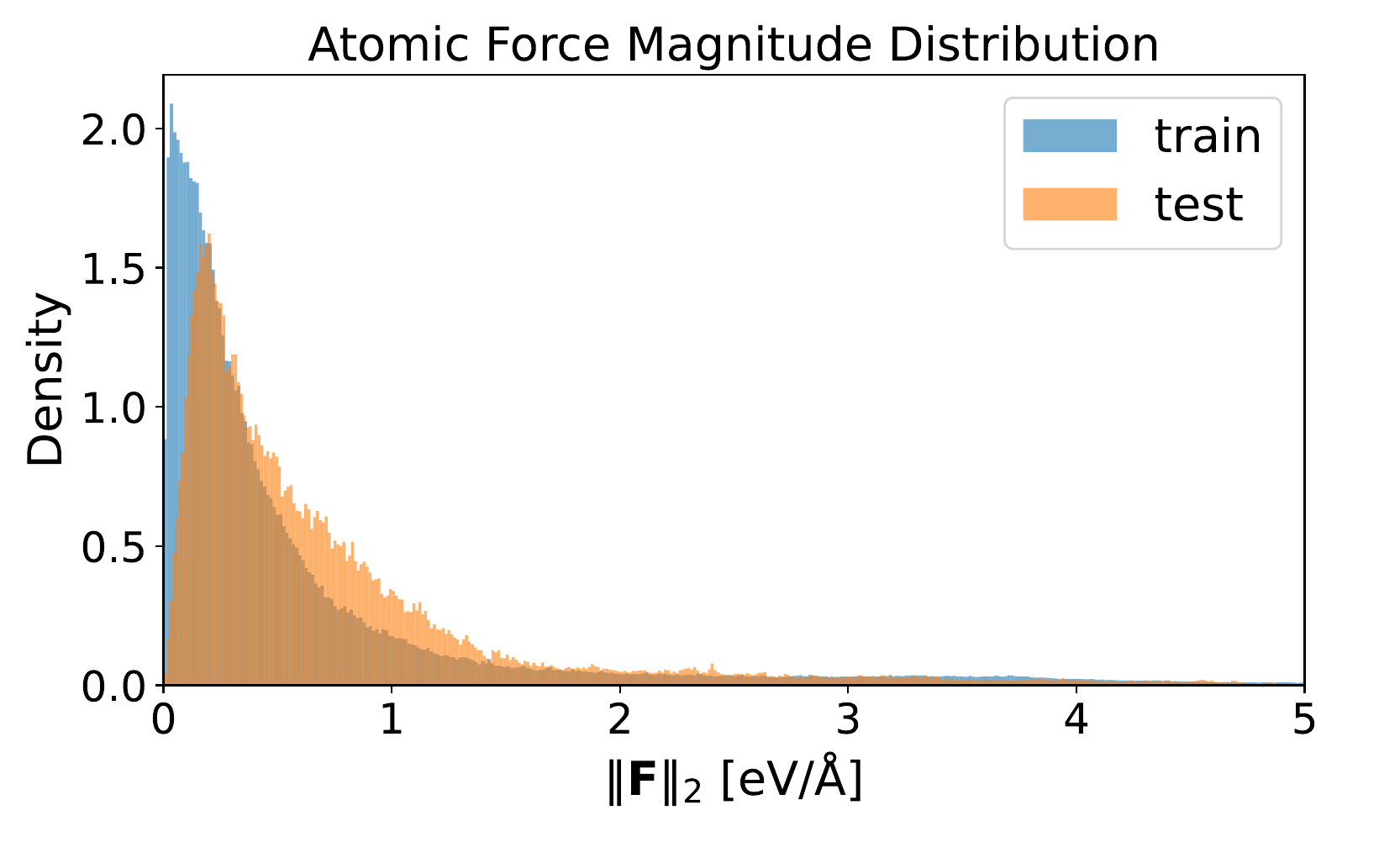}
        \caption{Forces per atom}
        \label{fig:data_without_forces}
    \end{subfigure}
    \caption{Distributions for energies, number of atoms, and mean force magnitudes (per atom and system) across training and test sets for the proposed dataset.
    } \vspace{-10pt}
\label{fig:histograms}
\end{figure}
\paragraph{Dataset statistics}
Each system is stored as a row in an HDF5 file containing atomic numbers, number of atoms, atomic positions, forces, formation energy, and unit cell. 
We also create corresponding XYZ files that are directly compatible with the MACE, UPET, MatRIS, and other frameworks.
\cref{tab:num_atoms_standard} summarises the atomic composition of the dataset, alongside the 1000 test sequences with \fivetimesish larger systems. \cref{fig:histograms} shows the distributions of formation energies, the number of atoms per system, and average L2-norms of forces (per system and per atom), respectively, for both training and test sets. As seen in Figure \ref{fig:energy_histogram}, the energy distributions differ between the training and test sets, whereas \cref{fig:force_per_system_histogram} reveals that many training systems are close to equilibrium (with low mean force magnitudes). Similar to the energies, certain force ranges are absent in the test data, further emphasising the difficulty of generalisation to unseen configurations.
\subsection{Combined DFT and ML approach}
To unlock the possibility of simulating MXenes under realistic conditions, we approximate DFT by modelling forces $\mathrm{f} \in \mathbb{R}^{N\times 3}$ and energies $E \in \mathbb{R}$ of MXene systems, with the atomic positions represented as the point set $\mathrm{system_{pos}}=\{\mathrm{pos}_i\}^N_{i=1} \in \mathbb{R}^{N \times 3}$ with $\mathrm{pos}_i \in \mathbb{R}^3$ being the $(x, y, z)$-coordinates of $\mathrm{atom}_i$, and $N$ the number of atoms in the system. Note that despite the two-dimensional periodic structure of MXenes, the atomic positions are still defined in three-dimensional Euclidean space $\mathbb{R}^3$. 

In atomistic modelling, translational invariance is typically enforced by using relative atomic positions, while rotational equivariance plays the central role in ensuring physically consistent force predictions.
When the coordinate system of $\mathrm{system_{pos}}$ is rotated, the forces $\mathrm{f}$, acting on the atoms, rotate accordingly, exhibiting the rotation equivariance of the $\mathrm{system}$.
That is, for a rotation-equivariant function $M_\mathrm{f}:~\mathbb{R}^{N\times 3} \rightarrow \mathbb{R}^{N\times 3}$ and any rotation $R\in \textup{SO}(3)$ represented as a matrix in $\mathbb{R}^{3\times 3}$,
\begin{equation}
\label{eq:equivariance}
    M_\mathrm{f}(\mathrm{system_{pos}} \, R) = M_\mathrm{f}(\mathrm{system_{pos}}) \, R \;.
\end{equation}
Simultaneously, this does not affect the energy of $\mathrm{system}$, \ie the energy is rotation-invariant in this context. That is, for a rotation-invariant function $M_E: \mathbb{R}^{N\times 3} \rightarrow \mathbb{R}
$,
\begin{equation}
\label{eq:invariance}
    M_E(\mathrm{system_{pos}} \, R) = M_E(\mathrm{system_{pos}}) \;. 
\end{equation}
There is a plethora of work demonstrating that incorporating rotation equivariance as an inductive bias into the model operating on point sets is useful \cite{anderson2019cormorant, ruhe2023clifford, equihypers}, including specifically for MLIPs \cite{liaoequiformer, passaro23a, aykent2025gotennet, fu2025learning}.


We select four such representative families of models, approximating forces and energy in accordance with \cref{eq:equivariance,eq:invariance}, respectively, and have demonstrated competitive performance in related molecular, materials, and catalysis applications:
EquiformerV2~\cite{Equiformerv2}, MACE~\cite{batatia2022mace}, and MatRIS, which integrate these SO$(3)$-symmetries \textit{by design}, and (U)PET, which learns these symmetries approximately from data augmentation.
The key architectural difference between EquiformerV2 and MACE is that MACE predicts forces by differentiating its energy output with respect to atomic positions using autograd, ensuring that forces are \textit{conservative} and consistent with the energy surface. 
In contrast, EquiformerV2 includes a dedicated force head and directly regresses forces as an independent output (\textit{non-conservative}), which can in some settings improve force accuracy but does not guarantee energy–force consistency. 
That is, MACE enforces physical consistency, while EquiformerV2 prioritises flexibility and performance in force prediction.
For MatRIS and (U)PET models, both conservative and non-conservative modes are available, which is valuable for our subsequent MD analysis and comparison.

In addition, while equivariance ensures robustness to coordinate transformations, it does not by itself guarantee reliable extrapolation to larger atomic systems, which we therefore investigate explicitly using the selected models in the following section.
\section{Experiments and Results}
We conduct experiments using two variants of EquiformerV2: the 30.8M-parameter model from the original paper \cite{Equiformerv2} (referred to as \textit{original}), and a smaller 4.8M-parameter model (referred to as \textit{small}). Among others, the main changes between \textit{original} and \textit{small} were reducing the number of layers from 8 to 5 and the number of attention heads from 8 to 4.  
For MACE, we use the standard model configuration comprising approximately $0.8$M parameters. In addition, we consider two foundation MACE models. The first foundation model, which we will call MACE \textit{PT omat} is trained on a combination of datasets, including OMAT~\cite{barroso_omat24}, OMOL~\cite{levine2025openmolecules2025omol25}, OC20~\cite{Chanussot2021} and MATPES~\cite{kaplan2025matpes}. It contains 6 heads, and we select the \textit{omat-pbe} head for task-specific fine-tuning, resulting in a model with approximately $6.2$M parameters~\cite{batatia2023foundation}. 
This head demonstrates state-of-the-art performance across inorganic, organic, and surface systems, and corresponds to the DFT level of theory that most closely matches our own. The second foundation model, which we call MACE \textit{PT matpes}, was trained on the MATPES dataset~\cite{kaplan2025matpes}, which contains only a single head and similarly, $0.8$M parameters. 
We also use the UPET~\cite{UPET-2026} models with conservative ($2.9$M) and non-conservative ($3.0$M) force-prediction mechanisms, as well as PET ($25.9$M) pretrained on the MAD-1.5 dataset~\cite{PET-MAD-1.5-2026}, denoted PET \emph{PT mad-s}.
For MatRIS, we use the standard model configuration with conservative forces, resulting in a model containing approximately $6.3$M parameters. We additionally investigate a non-conservative variant, which increases the parameter count slightly to approximately $6.4$M. Furthermore, we fine-tune two pretrained foundation models, each containing approximately $10.4$M parameters. The first model, denoted \textit{oam}, was pretrained on the OMat24 dataset and subsequently fine-tuned on the sAlex+MPTrj dataset. The second model, denoted \textit{mp}, was pretrained on the MPTrj dataset\footnote{Links to the foundation models can be found in \url{https://github.com/HPC-AI-Team/MatRIS/blob/main/matris/model/model.py}, in the  \texttt{load()} method.}.

\subsection{Training setup}
We first split the training set into training and validation sets with the $80/20$ ratio, aiming at having a similar distribution of system sizes (the validation split/indices are provided in the dataset; see also Table~\ref{tab:num_atoms_standard}).
For all the models, we utilise their official PyTorch~\cite{paszke2019pytorch} implementations\footnote{\url{https://github.com/atomicarchitects/equiformer_v2}, \url{https://github.com/ACEsuit/mace}, \url{https://github.com/lab-cosmo/upet}, and \url{https://github.com/HPC-AI-Team/MatRIS}}.

We train all the EquiformerV2 models end-to-end for 100 epochs, with batch size set to 1.
The learning rate is initially set to $4\cdot10^{-4}$ with a cosine annealing decay down to $4\cdot10^{-6}$ at the last epoch, with the AdamW optimiser. 
For MACE, we adopt the standard training hyperparameters, only modifying the loss function and decreasing the number of training epochs to $500$. 
With MatRIS, we follow the training procedure described in the paper as closely as possible.
Due to memory constraints associated with force prediction, training is performed using a batch size of 1. 
All models are trained for a maximum of 100 epochs.
At the same time, UPET models are trained for 300 epochs (except 100 epochs for finetuning the pretrained PET \textit{PT mad-s}), batch size 8, initial learning rate $10^{-4}$, and the remaining default metatrain hyperparameters.

By default, we use a mean absolute error (MAE) training objective
\begin{equation}
\label{eq:train_objective}
\begin{aligned}
\mathcal{L}_{\textup{\tiny total}} 
&:= 
    \lambda_E\, \underbrace{\frac{\sum_{s \in \mathcal{D}} |E_s^{\,\text{\tiny gt}} - E_s^{\,\text{\tiny pred}}|}{\sum_{s \in \mathcal{D}}N_s}}_{\mathcal{L}^\text{atom-weighted}_E} + \lambda_{\mathrm{f}}\,\underbrace{\frac{\sum_{s \in \mathcal{D}}\sum_{i=1}^{N_s} \sum_{j=1}^3 {|\mathrm{f}^{\,\text{\tiny gt}}_{sij} - \mathrm{f}^{\,\text{\tiny pred}}_{sij}|}}{\sum_{s \in \mathcal{D}} 3N_\text{s}}}_{\mathcal{L}_\mathrm{f}}\; , 
\end{aligned}
\end{equation}
for a system $s$ with $N_s$ atoms, given a dataset $\mathcal{D}$, where $\mathrm{gt}$ and $\mathrm{pred}$ are respectively ground-truth and predicted quantities. 
For UPET, we use its standard metatrain pipeline, where the per-atom energy error is optimised uniformly over the total number of structures, $|\mathcal{D}|$, \ie $\mathcal{L}_E:=\frac{1}{|\mathcal{D}|}\sum_{s \in \mathcal{D}} \frac{|E_s^{\,\text{\tiny gt}} - E_s^{\,\text{\tiny pred}}|}{N_s}$ is used for the energy term in \cref{eq:train_objective}, while $\mathcal{L}_\text{f}$ remains the same.
We experimentally find that $\lambda_E = 1,~\lambda_{\mathrm{f}}=25$ works best for EquiformerV2, while $\lambda_E = 1,~\lambda_{\mathrm{f}}=1$ for MACE, $\lambda_E = 1,~\lambda_{\mathrm{f}}=5$ for UPET, and $\lambda_E = 5,~\lambda_{\mathrm{f}}=5$ for MatRIS.
We train all models until convergence to the best validation performance attainable within our computational budget. 
For our experiments, we use NVIDIA A100 (40 GB), A100 FAT (80 GB), and H100 (141 GB).

\begin{table*}[t]
\centering
\scriptsize
\setlength{\tabcolsep}{3.5pt}
\renewcommand{\arraystretch}{1.2}

\makebox[\textwidth][c]{%
\begin{tabular}{l c c c c c}
\toprule
\multirow[b]{2}{*}{\textbf{Model}} &
\multirow[b]{2}{*}{\makecell[c]{\textbf{Training}\\\textbf{time per run}\\\textbf{(GPU h)}}} &
\multicolumn{2}{c}{\textbf{Test original, MAE}} &
\multicolumn{2}{c}{\textbf{Test larger systems (\fivetimesish), MAE}} \\
\cmidrule(lr){3-4}\cmidrule(lr){5-6}
& &
\makecell{Energy/atom\\(meV) $\downarrow$} &
\makecell{Force\\(meV/\AA) $\downarrow$} &
\makecell{Energy/atom\\(meV) $\downarrow$} &
\makecell{Force\\(meV/\AA) $\downarrow$} \\
\midrule

EquiformerV2 \textit{Original} (30.8M) &
253.6 &
3.7 $\pm$ 0.2 (3.5) &
16.0 $\pm$ 0.2 (15.7) &
5.3 $\pm$ 0.2 (5.0) &
46.0 $\pm$ 0.2 (45.9) \\

EquiformerV2 \textit{Small} (4.8M) &
141.3 &
3.6 $\pm$ 0.1 (3.5) &
15.9 $\pm$ 0.7 (15.0) &
\makecell{5.0 $\pm$ 0.2 (4.8)} &
\makecell{44.4 $\pm$ 0.4 (44.0)} \\

\midrule

MACE (0.8M) &
234.3 &
2.1 $\pm$ 0.1 (2.0) &
15.5 $\pm$ 0.8 (14.8) &
\makecell{\underline{1.4} $\pm$ 0.1 (1.3)} &
\makecell{42.6 $\pm$ 0.2 (42.3)} \\

MACE \textit{PT omat} (6.2M) &
108.6* & 
1.8 $\pm$ 0.2 (1.7) &
14.6 $\pm$ 4.1 (11.7) &
\makecell{1.9 $\pm$ 0.3 (1.6)} &
\makecell{47.4 $\pm$ 4.6 (43.5)} \\

MACE\textit{ PT matpes} (0.8M) &
60.4* &
\underline{1.7} $\pm$ 0.1 (1.6) &
\underline{12.2} $\pm$ 0.2 (11.9) &
\makecell{1.8 $\pm$ 0.2 (1.6)} &
\makecell{43.9 $\pm$ 1.2 (42.8)} \\

\midrule
UPET cons. (2.9M) &
153.5 &
\makecell{ 1.3 $\pm$ 0.1 (1.2)} &
\makecell{20.6 $\pm$ 0.2 (20.4)} &
\makecell{2.5 $\pm$ 0.2 (2.3)} &
\makecell{57.8 $\pm$ 1.2 (56.0)} \\

UPET non-cons. (3.0M) &
72.1 &
\makecell{2.5 $\pm$ 0.1 (2.4)} &
\makecell{26.9 $\pm$ 1.6 (24.7)} &
\makecell{2.4 $\pm$ 0.5 (1.9)} &
\makecell{49.4 $\pm$ 1.4 (47.7)} \\

PET \textit{PT mad-s} (25.9M) &
36.5 &
\makecell{\textbf{1.1} $\pm$ 0.0 (1.0)} &
\makecell{\textbf{10.3} $\pm$ 0.0 (10.3)} &
\makecell{\textbf{1.3} $\pm$ 0.0 (1.3)} &
\makecell{\textbf{36.3} $\pm$ 0.1 (36.2)}  \\

\midrule
MatRIS cons. (6.3M) &
 576.0 &
\makecell{ 4.8 $\pm$ 0.4 (4.4)} &
\makecell{ 23.5 $\pm$ 1.9 (21.6)} &
\makecell{ 3.3 $\pm$ 0.3 (3.1)} &
\makecell{ 46.6 $\pm$ 1.4 (45.2)} \\

MatRIS non-cons. (6.4M) &
 576.0 &
\makecell{ 5.8 $\pm$ 0.9 (4.8)} &
\makecell{ 22.2 $\pm$ 1.0 (21.0)} &
\makecell{ 5.1 $\pm$ 1.8 (3.9)} &
\makecell{ 48.7 $\pm$ 1.2 (47.4)} \\

MatRIS \textit{PT mp} (10.4M) &
 576.0 &
\makecell{ 7.0 $\pm$ 1.2 (5.9)} &
\makecell{35.7 $\pm$ 2.3 (33.7)} &
\makecell{5.0 $\pm$ 2.6 (3.4)} &
\makecell{44.8 $\pm$ 2.4 (42.3)} \\

MatRIS \textit{PT oam} (10.4M) &
 576.0 &
\makecell{5.2 $\pm$ 2.7 (3.4)} &
\makecell{18.4 $\pm$ 1.0 (17.6)} &
\makecell{3.7 $\pm$ 2.1 (2.0)} &
\makecell{\underline{40.1} $\pm$ 0.9 (39.1)} \\

\bottomrule
\end{tabular}
}

\caption{
Test performance of equivariant MLIPs. 
Evaluation of MLIP models trained on the proposed dataset and evaluated on the original test set and test set with larger systems (\fivetimesish).
Reported are the MAE mean and standard deviation over three runs, with the best run result in parentheses.
The best results overall are highlighted in bold, with the second best underlined.
* indicates training on H100 instead of A100 (note that for EquiformerV2 and MACE, we used A100 FAT).
\vspace{-10pt}
}
\label{tab:cross_evaluation}
\end{table*}
\vspace{-10pt}
\subsection{Results}
\vspace{-5pt}
\cref{tab:cross_evaluation} summarises test results of all the models on the proposed dataset (\textbf{Test original} and, for the generalisation test, \textbf{Test larger systems}), where energy MAE refers to $\mathcal{L}_E$, and force MAE to $\mathcal{L}_\mathrm{f}$ from \cref{eq:train_objective}, which corresponds to the standard evaluation metrics in the literature. 
Similar to ~\cite{batatia2022mace} and others, we also report root mean square error (RMSE) scores in~Appendix~\ref{app:quantitative}.
For each run, the evaluation on the test set is conducted using the models with the best validation performance. 
We first note that reducing EquiformerV2 capacity does not degrade its performance: EquiformerV2 \textit{small} achieves comparable MAE and noticeably lower RMSE than the larger EquiformerV2 \textit{original}, indicating improved robustness to outliers despite a substantially smaller parameter count. 
Secondly, MACE models consistently outperform both EquiformerV2 variants in energy accuracy and achieve comparable or better force MAE, with the smallest 0.8M-parameter MACE already surpassing the larger Equiformer models. 
Fine-tuning pretrained MACE foundation models yields further improvements across all metrics, with MACE \textit{PT matpes} achieving the best balance between energy and force accuracy among the MACE variants.
The PET \textit{PT mad-s} model achieves the lowest MAE overall on the original test set, reaching $1.1$ meV/atom for energies and $10.3$ meV/Å for forces, while also exhibiting strong performance on larger systems. 
When evaluating generalisation to larger systems, most models show an increase in force MAE, whereas energy MAE remains comparatively stable. 
PET \textit{PT mad-s} and pretrained MACE models show the smallest degradation, indicating better generalisation to the input system size. 

Comparing conservative and non-conservative force formulations reveals a consistent advantage for conservative models in terms of energy accuracy and physical consistency.
For UPET, enforcing energy–force consistency substantially improves both energy accuracy and force accuracy on the original test set ($1.3$ vs. $2.5$ meV/atom and $20.6$ vs. $26.9$ meV/Å), although both variants show notable degradation on larger systems. 
A similar trend is observed for MatRIS, where the conservative model outperforms the non-conservative version on the original test set for energy MAE, with improved energy generalisation to larger systems. 
Overall, conservative force prediction systematically improves energy accuracy and physical consistency, while force MAE remains comparable to non-conservative formulations and shows improvements primarily on in-distribution data for selected models.
A more detailed analysis is provided in~Appendix~\ref{app:quantitative}.
\begin{wraptable}{L}{0.5\textwidth}
\centering
\scriptsize
\setlength{\tabcolsep}{6pt}
\renewcommand{\arraystretch}{1.2}
\begin{tabular}{@{} l cc}
\toprule
\multirow{2}{*}{\textbf{Model}} &
\multicolumn{2}{c}{\textbf{median system (96 atoms)}} \\
\cmidrule(lr){2-3}
& \textbf{CPU} & \textbf{GPU} \\
\midrule

DFT & 890 s & --  \\

\midrule

EquiformerV2 \textit{Original} (30.8M)
& 0.823 s & 0.070 s \\

EquiformerV2 \textit{Small} (4.8M)
& 0.300 s & 0.040 s \\

\midrule 

MACE (0.8M)
& 0.209 s & 0.028 s  \\

MACE \textit{PT omat} (6.2M)
& 0.385 s & 0.035 s \\

\midrule
UPET cons. (2.9M) & 0.947 s & 0.022 s \\

UPET non-cons. (3.0M) & 0.512 s & 0.012 s \\

PET \textit{PT mad-s} (25.9M) & 0.701 s & 0.024 s \\

\midrule
MatRIS cons. (6.3M) & 1.830 s & 0.120 s \\

MatRIS non-cons. (6.4M) & 0.874 s & 0.058 s \\

MatRIS \textit{PT oam} (10.4M) & 4.210 s & 0.309 s \\

MatRIS \textit{PT mp} (10.4M) & 4.413 s & 0.310 s \\

\bottomrule
\end{tabular}
\caption{
Comparison of inference times for DFT and ML interatomic potentials on CPU (Intel Xeon Gold 6130 @ 2.10\,GHz) and GPU (A100 FAT).
Times are reported for a median-sized configuration in the test set (96 atoms).\vspace{-12pt}
}
\label{tab:speed_comparison}
\end{wraptable}
\begin{figure}
\centering
\includegraphics[width=\textwidth]{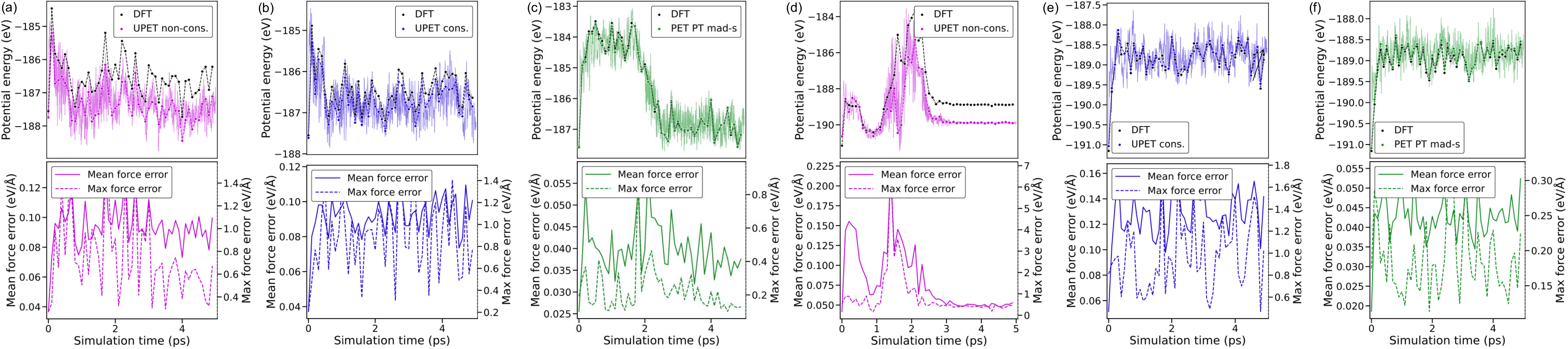}
\caption{Molecular dynamics simulations of (a-c) CO$_2$ on a fully OH-terminated MXene and (d-f) HCOOH on a fully O-terminated MXene at 300 K, comparing the performance of the (a,d) UPET non-cons., (b,e) UPET cons., and (c,f) PET \textit{PT mad-s} models. Each simulation consists of 10 000 time steps with a step size of 0.5 fs. At every 200th step, single-point DFT calculations are performed; top panels show the DFT and ML potential energies along the trajectories (defined with respect to elemental reference energies according to Eq.~\ref{eq:formation-energy}), while bottom panels report the mean and maximum per-atom force L2 errors relative to DFT.}\vspace{-12pt}
\label{fig:MD comparison_PET}
\end{figure}
\vspace{-10pt}
\paragraph{Speed comparison} We display the comparison between the inference time of the trained models and DFT in \cref{tab:speed_comparison}. 
We note that even on a CPU, all the trained models run significantly faster than DFT.
In particular, two of the more accurate models, MACE and PET, achieve a computational workflow acceleration of 4,258- and 1,270-fold, respectively.
\vspace{-10pt}
\paragraph{Qualitative assessment}
To complement the quantitative benchmarks, we perform a qualitative analysis to identify model weaknesses not evident from standard test-set metrics. 
Pronounced differences emerge when the models are used for molecular dynamics (MD) simulations of CO$_2$ and HCOOH adsorbed on Ti$_2$C MXenes, as exemplified in Figure.~\ref{fig:MD comparison_PET}. 

For the adsorbed CO$_2$, the non-conservative and conservative UPET models have similar force errors, with a slightly larger energy mismatch for the non-conservative model. 
For the HCOOH adsorption, the errors for the non-conservative model are more severe, particularly in forces, whereas the conservative model performs similarly to that of CO$_2$. 
The same trends are seen when comparing non-conservative and conservative MatRIS (Figure~\ref{fig:MD comparison_matris}), as well as for the conservative MACE and non-conservative EquiformerV2 (Figure~\ref{fig:MD comparison_mace-equi}), pointing 
towards the importance of conservative force formulations for machine learning potentials for stable MD simulations. 
Among the conservative models, MatRIS gives the largest errors, specifically for the system with the adsorbed CO$_2$. 

The fine-tuned PET PT \emph{mad-s} model, which performed best in the quantitative evaluation, also yields the smallest deviations in the qualitative MD assessment (Figure~\ref{fig:MD comparison_PET}c,f), particularly with respect to forces, followed by the fine-tuned MACE \emph{PT matpes} and \emph{omat} (Figure~\ref{fig:MD CO2 comparison} and \ref{fig:MD HCOOH comparison}). We also note that analysis based on the radial distribution function captures the large discrepancy observed in the qualitative MD analysis, but does not distinguish the performance of the conservative models when comparing the PET-based models (Figure~\ref{fig:rdfs pets}).
\vspace{-5pt}
\section{Conclusions}
\vspace{-7pt}
We have generated a dataset for catalysis on Ti$_2$CT$_y$ MXenes, called \datasetname, comprising a total of 60,000 DFT calculations --- 50,000 for training and 10,000 for testing. In addition, we construct a dataset of 1,000 larger systems to explicitly evaluate model generalisation to increased system sizes. The construction of \datasetname is motivated by the limited reliability of existing foundation models when applied to MXene-based catalytic systems. 
\vspace{-8pt}
\paragraph{Limitations} 
Due to computational limitations, the hyperparameter search we have conducted is non-exhaustive.
In the DFT training data, the main limitation concerns the molecular systems included. 
The chosen molecules represent prototypical catalytic reactions, providing 
chemically meaningful examples for MXene interactions. 
Future work could extend the molecular scope to additional reactants and intermediates relevant to other catalytic processes.

\section*{Impact Statement}
This work accelerates the analysis of catalysis on 2D MXenes while maintaining robustness through the trained model. Faster evaluations of MXene catalysis open opportunities for identifying more efficient catalysts, supporting efforts to reduce the carbon footprint of the chemical industry. In addition, the generated dataset provides a resource for the community to develop and benchmark models for catalysis on Ti$_2$CT$_y$ MXenes and related materials.

\section*{Acknowledgments}
This work was supported by the Wallenberg AI, Autonomous Systems and Software Program (WASP) and the Wallenberg Initiative Materials Science for Sustainability (WISE), by the Swedish Research Council through a
grant for the project Uncertainty-Aware Transformers for Regression Tasks in Computer Vision (2022-04266), and the strategic research environment ELLIIT. The computations were enabled by resources provided by the National Academic Infrastructure for Supercomputing in Sweden (NAISS) partially funded by the Swedish Research Council through grant agreement no. 2022-06725, and by the Berzelius resource provided by the
Knut and Alice Wallenberg Foundation at the National Supercomputer Centre.

\bibliographystyle{abbrvnat}
\bibliography{references} 

@article{batatia2022mace,
  title={{MACE: Higher order equivariant message passing neural networks for fast and accurate force fields}},
  author={Batatia, Ilyes and Kovacs, David P and Simm, Gregor and Ortner, Christoph and Cs{\'a}nyi, G{\'a}bor},
  journal={Advances in neural information processing systems},
  volume={35},
  pages={11423--11436},
  year={2022}
}

@article{ase-paper,
  author={Ask Hjorth Larsen and Jens Jørgen Mortensen and Jakob Blomqvist and Ivano E Castelli and Rune Christensen and Marcin
Dułak and Jesper Friis and Michael N Groves and Bjørk Hammer and Cory Hargus and Eric D Hermes and Paul C Jennings and Peter
Bjerre Jensen and James Kermode and John R Kitchin and Esben Leonhard Kolsbjerg and Joseph Kubal and Kristen
Kaasbjerg and Steen Lysgaard and Jón Bergmann Maronsson and Tristan Maxson and Thomas Olsen and Lars Pastewka and Andrew
Peterson and Carsten Rostgaard and Jakob Schiøtz and Ole Schütt and Mikkel Strange and Kristian S Thygesen and Tejs
Vegge and Lasse Vilhelmsen and Michael Walter and Zhenhua Zeng and Karsten W Jacobsen},
  title={The atomic simulation environment—a Python library for working with atoms},
  journal={J. Phys. Condens. Matter},
  volume={29},
  number={27},
  pages={273002},
  url={http://stacks.iop.org/0953-8984/29/i=27/a=273002},
  year={2017},
}

@article{Blochl1994,
	title = {Projector augmented-wave method},
	volume = {50},
	issn = {0163-1829, 1095-3795},
	url = {https://link.aps.org/doi/10.1103/PhysRevB.50.17953},
	doi = {10.1103/PhysRevB.50.17953},
	number = {24},
	journal = {Phys. Rev. B},
	author = {Blöchl, P. E.},
	month = dec,
	year = {1994},
	pages = {17953--17979},
}

@article{Chanussot2021,
	title = {Open catalyst 2020 ({OC20}) dataset and community challenges},
	volume = {11},
	copyright = {https://doi.org/10.15223/policy-029},
	issn = {2155-5435, 2155-5435},
	url = {https://pubs.acs.org/doi/10.1021/acscatal.0c04525},
	doi = {10.1021/acscatal.0c04525},
	number = {10},
	journal = {ACS Catal.},
	author = {Chanussot, Lowik and Das, Abhishek and Goyal, Siddharth and Lavril, Thibaut and Shuaibi, Muhammed and Riviere, Morgane and Tran, Kevin and Heras-Domingo, Javier and Ho, Caleb and Hu, Weihua and Palizhati, Aini and Sriram, Anuroop and Wood, Brandon and Yoon, Junwoong and Parikh, Devi and Zitnick, C. Lawrence and Ulissi, Zachary},
	month = may,
	year = {2021},
	pages = {6059--6072},
}

@article{Chen2025,
	title = {A density functional benchmark for dehydrogenation and dehalogenation reactions on coinage metal surfaces},
	volume = {26},
	issn = {1439-4235, 1439-7641},
	url = {https://chemistry-europe.onlinelibrary.wiley.com/doi/10.1002/cphc.202400865},
	doi = {10.1002/cphc.202400865},
	number = {1},
	journal = {ChemPhysChem},
	author = {Chen, Lin and Rosen, Johanna and Björk, Jonas},
	month = jan,
	year = {2025},
	pages = {e202400865},
}

@article{Dion2004,
	title = {Van der Waals density functional for general geometries},
	volume = {92},
	issn = {0031-9007, 1079-7114},
	url = {https://link.aps.org/doi/10.1103/PhysRevLett.92.246401},
	doi = {10.1103/PhysRevLett.92.246401},
	language = {en},
	number = {24},
	journal = {Phys. Rev. Lett.},
	author = {Dion, M. and Rydberg, H. and Schröder, E. and Langreth, D. C. and Lundqvist, B. I.},
	month = jun,
	year = {2004},
	pages = {246401},
}

@article{Guo2025,
	title = {Machine learning-guided discovery of thermodynamically stable single-atom catalysts on functionalized {MXenes} for enhanced oxygen reduction and evolution reactions},
	volume = {13},
	issn = {2050-7488, 2050-7496},
	url = {https://xlink.rsc.org/?DOI=D5TA02929E},
	doi = {10.1039/D5TA02929E},
	number = {28},
	urldate = {2025-10-07},
	journal = {J. Mater. Chem. A},
	author = {Guo, Hengquan and Lee, Seung Geol},
	year = {2025},
	pages = {22730--22744},
}

@article{Hamada2014,
	title = {van der Waals density functional made accurate},
	volume = {89},
	issn = {1098-0121, 1550-235X},
	url = {https://link.aps.org/doi/10.1103/PhysRevB.89.121103},
	doi = {10.1103/PhysRevB.89.121103},
	number = {12},
	journal = {Phys. Rev. B},
	author = {Hamada, Ikutaro},
	year = {2014},
	pages = {121103},
}

@article{Hohenberg1964,
	title = {Inhomogeneous electron gas},
	volume = {136},
	issn = {0031-899X},
	url = {https://link.aps.org/doi/10.1103/PhysRev.136.B864},
	doi = {10.1103/PhysRev.136.B864},
	number = {3B},
	urldate = {2025-09-13},
	journal = {Phys. Rev.},
	author = {Hohenberg, P. and Kohn, W.},
	year = {1964},
	pages = {B864--B871},
}

@article{Kohn1965,
	title = {Self-consistent equations including exchange and correlation effects},
	volume = {140},
	issn = {0031-899X},
	url = {https://link.aps.org/doi/10.1103/PhysRev.140.A1133},
	doi = {10.1103/PhysRev.140.A1133},
	number = {4A},
	journal = {Phys. Rev.},
	author = {Kohn, W. and Sham, L. J.},
	year = {1965},
	pages = {A1133--A1138},
}

@article{Kresse1996,
	title = {Efficient iterative schemes for \textit{ab initio} total-energy calculations using a plane-wave basis set},
	volume = {54},
	issn = {0163-1829, 1095-3795},
	url = {https://link.aps.org/doi/10.1103/PhysRevB.54.11169},
	doi = {10.1103/PhysRevB.54.11169},
	number = {16},
	journal = {Phys. Rev. B},
	author = {Kresse, G. and Furthmüller, J.},
	month = oct,
	year = {1996},
	pages = {11169--11186},
}

@article{Kresse1999,
	title = {From ultrasoft pseudopotentials to the projector augmented-wave method},
	volume = {59},
	issn = {0163-1829, 1095-3795},
	url = {https://link.aps.org/doi/10.1103/PhysRevB.59.1758},
	doi = {10.1103/PhysRevB.59.1758},
	language = {en},
	number = {3},
	journal = {Phys. Rev. B},
	author = {Kresse, G. and Joubert, D.},
	month = jan,
	year = {1999},
	pages = {1758--1775},
}

@article{Lin2025,
	title = {Machine learning accelerated screening advanced single-atom anchored {MXenes} electrocatalyst for nitrogen fixation},
	volume = {15},
	copyright = {https://doi.org/10.15223/policy-029},
	issn = {2155-5435, 2155-5435},
	url = {https://pubs.acs.org/doi/10.1021/acscatal.4c06914},
	doi = {10.1021/acscatal.4c06914},
	language = {en},
	number = {15},
	journal = {ACS Catal.},
	author = {Lin, Gaobo and Guo, Teng and Lin, Wenwen and Fan, Haoan and Guo, Lei and Zhang, Zhenyu and Li, Bolong and Wang, Jianghao and Ji, Huiping and Song, Weiyu and Fu, Jie},
	month = aug,
	year = {2025},
	pages = {13534--13548},
}

@article{Niu2020,
	title = {C–H activation of light alkanes on MXenes predicted by hydrogen affinity},
	volume = {22},
	issn = {1463-9076, 1463-9084},
	url = {https://xlink.rsc.org/?DOI=D0CP02471F},
	doi = {10.1039/D0CP02471F},
	number = {33},
	journal = {Phys. Chem. Chem. Phys.},
	author = {Niu, Kaifeng and Chi, Lifeng and Rosen, Johanna and Bj\"ork, Jonas},
	year = {2020},
	pages = {18622--18630},
}

@article{Niu2021,
	title = {Structure-activity correlation of {Ti}$_{\textrm{2}}${CT}$_{\textrm{2}}$ MXenes for C–H activation},
	volume = {33},
	issn = {0953-8984, 1361-648X},
	url = {https://iopscience.iop.org/article/10.1088/1361-648X/abe8a1},
	doi = {10.1088/1361-648X/abe8a1},
	number = {23},
	journal = {J. Phys. Condens. Matter},
	author = {Niu, Kaifeng and Chi, Lifeng and Rosen, Johanna and Bj\"ork, Jonas},
	month = jun,
	year = {2021},
	pages = {235201},
}

@article{Niu2025,
	title = {First-principles exploration of {Sc}- and {Y}-based {MXenes} with halogen terminations},
	volume = {9},
	issn = {2397-7132},
	url = {https://www.nature.com/articles/s41699-025-00589-7},
	doi = {10.1038/s41699-025-00589-7},
	language = {en},
	number = {1},
	urldate = {2025-09-09},
	journal = {npj 2D Mater. Appl.},
	author = {Niu, Kaifeng and Björk, Jonas and Rosen, Johanna},
	year = {2025},
	pages = {69},
}

@article{Parui2022,
	title = {Selective reduction of {CO}$_{\textrm{2}}$ on {Ti}$_{\textrm{2}}${C}({OH})$_{\textrm{2}}$ {MXene} through spontaneous crossing of transition states},
	volume = {14},
	issn = {1944-8244, 1944-8252},
	url = {https://pubs.acs.org/doi/10.1021/acsami.2c10213},
	doi = {10.1021/acsami.2c10213},
	number = {36},
	journal = {ACS Appl. Mater. Interfaces},
	author = {Parui, Arko and Srivastava, Pooja and Singh, Abhishek Kumar},
	year = {2022},
	pages = {40913--40920},
}

@article{Tran2019,
	title = {Nonlocal van der {Waals} functionals for solids: {Choosing} an appropriate one},
	volume = {3},
	issn = {2475-9953},
	shorttitle = {Nonlocal van der {Waals} functionals for solids},
	url = {https://link.aps.org/doi/10.1103/PhysRevMaterials.3.063602},
	doi = {10.1103/PhysRevMaterials.3.063602},
	number = {6},
	journal = {Phys. Rev. Materials},
	author = {Tran, Fabien and Kalantari, Leila and Traoré, Boubacar and Rocquefelte, Xavier and Blaha, Peter},
	month = jun,
	year = {2019},
	pages = {063602},
}

@article{Tran2023,
  author = {Tran, Richard and Lan, Janice and Shuaibi, Muhammed and Wood, Brandon M. and Goyal, Siddharth and Das, Abhishek and Heras-Domingo, Javier and Kolluru, Adeesh and Rizvi, Ammar and Shoghi, Nima and Sriram, Anuroop and Therrien, Félix and Abed, Jehad and Voznyy, Oleksandr and Sargent, Edward H. and Ulissi, Zachary and Zitnick, C. Lawrence},
  title = {The Open Catalyst 2022 (OC22) Dataset and Challenges for Oxide Electrocatalysts},
  journal = {ACS Catal.},
  volume = {13},
  number = {5},
  pages = {3066--3084},
  year = {2023},
  doi = {10.1021/acscatal.2c05426},
  url = {https://doi.org/10.1021/acscatal.2c05426}
}

@article{Equiformerv2,
  title={{EquiformerV2: Improved Equivariant Transformer for Scaling to Higher-Degree Representations}},
  author={Liao, Yi-Lun and Wood, Brandon and Das, Abhishek and Smidt, Tess},
  journal={The Twelfth International Conference on Learning Representations},
  year={2024}
}

@article{MLIP_DFT_1,
  title={Machine learning interatomic potentials as emerging tools for materials science},
  author={Deringer, Volker L and Caro, Miguel A and Cs{\'a}nyi, G{\'a}bor},
  journal={Adv. Mater.},
  volume={31},
  number={46},
  pages={1902765},
  year={2019},
  publisher={Wiley Online Library}
}

@article{MLIP_DFT_2,
  title={A universal graph deep learning interatomic potential for the periodic table},
  author={Chen, Chi and Ong, Shyue Ping},
  journal={Nat. Comput. Sci.},
  volume={2},
  number={11},
  pages={718--728},
  year={2022},
  publisher={Nature Publishing Group US New York}
}

@article{mattersim,
  title={{MatterSim: A deep learning atomistic model across elements, temperatures and pressures}},
  author={Yang, Han and Hu, Chenxi and Zhou, Yichi and Liu, Xixian and Shi, Yu and Li, Jielan and Li, Guanzhi and Chen, Zekun and Chen, Shuizhou and Zeni, Claudio and others},
  journal={arXiv preprint arXiv:2405.04967},
  year={2024}
}

@article{batzner20223,
  title={E (3)-equivariant graph neural networks for data-efficient and accurate interatomic potentials},
  author={Batzner, Simon and Musaelian, Albert and Sun, Lixin and Geiger, Mario and Mailoa, Jonathan P and Kornbluth, Mordechai and Molinari, Nicola and Smidt, Tess E and Kozinsky, Boris},
  journal={Nat. Commun.},
  volume={13},
  number={1},
  pages={2453},
  year={2022},
  publisher={Nature Publishing Group UK London}
}

@article{ruhe2023clifford,
title={{Clifford Group Equivariant Neural Networks}},
author={Ruhe, David and Brandstetter, Johannes and Forr{\'e}, Patrick},
journal={Thirty-seventh Conference on Neural Information Processing Systems},
year={2023},
url={https://openreview.net/forum?id=n84bzMrGUD}
}

@article{anderson2019cormorant,
  title={Cormorant: Covariant molecular neural networks},
  author={Anderson, Brandon and Hy, Truong Son and Kondor, Risi},
  journal={Advances in Neural Information Processing Systems},
  pages={14510--14519},
  year={2019}
}

@article{paszke2019pytorch,
  title={PyTorch: An imperative style, high-performance deep learning library},
  author={Paszke, Adam and Gross, Sam and Massa, Francisco and Lerer, Adam and Bradbury, James and Chanan, Gregory and Killeen, Trevor and Lin, Zeming and Gimelshein, Natalia and Antiga, Luca and others},
  journal={Advances in Neural Information Processing Systems},
  pages={8024--8035},
  year={2019}
}

@article{bronstein2017geometric,
  title={Geometric deep learning: going beyond Euclidean data},
  author={Bronstein, Michael M and Bruna, Joan and LeCun, Yann and Szlam, Arthur and Vandergheynst, Pierre},
  journal={IEEE Signal Processing Magazine},
  volume={34},
  number={4},
  pages={18--42},
  year={2017},
  publisher={IEEE}
}

@article{weiler20183d,
  title={{3D steerable {CNN}s: Learning rotationally equivariant features in volumetric data}},
  author={Weiler, Maurice and Geiger, Mario and Welling, Max and Boomsma, Wouter and Cohen, Taco S},
  journal={Advances in Neural Information Processing Systems},
  pages={10381--10392},
  year={2018}
}

@article{fuchs2020se3,
 author = {Fuchs, Fabian and Worrall, Daniel and Fischer, Volker and Welling, Max},
 journal = {Advances in Neural Information Processing Systems},
 editor = {H. Larochelle and M. Ranzato and R. Hadsell and M. F. Balcan and H. Lin},
 pages = {1970--1981},
 publisher = {Curran Associates, Inc.},
 title = {SE(3)-Transformers: 3D Roto-Translation Equivariant Attention Networks},
 volume = {33},
 year = {2020}
}

@article{van1995vision,
  title={Vision and Lie's approach to invariance},
  author={Van Gool, Luc and Moons, Theo and Pauwels, Eric and Oosterlinck, Andr{\'e}},
  journal={Image and vision computing},
  volume={13},
  number={4},
  pages={259--277},
  year={1995},
  publisher={Elsevier}
}

@article{bronstein2021geometric,
  title={{Geometric deep learning: Grids, groups, graphs, geodesics, and gauges}},
  author={Bronstein, Michael M and Bruna, Joan and Cohen, Taco and Veli{\v{c}}kovi{\'c}, Petar},
  journal={arXiv preprint arXiv:2104.13478},
  year={2021}
}

@article{liaoequiformer,
  title={{Equiformer: Equivariant Graph Attention Transformer for 3D Atomistic Graphs}},
  author={Liao, Yi-Lun and Smidt, Tess},
  journal={The Eleventh International Conference on Learning Representations},
  year={2023}
}

@article{tetrasphere,
    author    = {Melnyk, Pavlo and Robinson, Andreas and Felsberg, Michael and Wadenb\"ack, M\r{a}rten},
    title     = {{TetraSphere: A Neural Descriptor for O(3)-Invariant Point Cloud Analysis}},
    journal = {Proceedings of the IEEE/CVF Conference on Computer Vision and Pattern Recognition (CVPR)},
    month     = {6},
    year      = {2024},
    pages     = {5620-5630},
}

@article{equihypers,
  title = 	 {{O$n$ Learning Deep O($n$)-Equivariant Hyperspheres}},
  author =       {Melnyk, Pavlo and Felsberg, Michael and Wadenb\"{a}ck, M{\aa}rten and Robinson, Andreas and Le, Cuong},
  journal = 	 {Proceedings of the 41st International Conference on Machine Learning},
  pages = 	 {35324--35339},
  year = 	 {2024},
  volume = 	 {235},
  series = 	 {Proceedings of Machine Learning Research},
  month = 	 {7},
  publisher =    {PMLR},
  url = 	 {https://proceedings.mlr.press/v235/melnyk24a.html},
}

@article{weiler2023equivariant,
  title={Equivariant and coordinate independent convolutional networks},
  author={Weiler, Maurice and Forr{\'e}, Patrick and Verlinde, Erik and Welling, Max},
  journal={A Gauge Field Theory of Neural Networks},
  pages={110},
  year={2023},
  publisher={World Scientific Singapore}
}

@article{esteves17_learn_so_equiv_repres_with_spher_cnns,
  author = {Esteves, Carlos and Allen-Blanchette, Christine and Makadia, Ameesh and Daniilidis, Kostas},
  title = {{Learning SO(3) Equivariant Representations With Spherical CNNs}},
  journal = {CoRR},
  year = {2017},
  url = {http://arxiv.org/abs/1711.06721},
  archivePrefix = {arXiv},
  eprint = {1711.06721},
  primaryClass = {cs.CV},
}

@article{aykent2025gotennet,
  title={{Gotennet: Rethinking efficient 3d equivariant graph neural networks}},
  author={Aykent, Sarp and Xia, Tian},
  journal={The Thirteenth International Conference on Learning Representations},
  year={2025}
}

@article{passaro23a,
  title = 	 {Reducing {SO}(3) Convolutions to {SO}(2) for Efficient Equivariant {GNN}s},
  author =       {Passaro, Saro and Zitnick, C. Lawrence},
  journal = 	 {Proceedings of the 40th International Conference on Machine Learning},
  pages = 	 {27420--27438},
  year = 	 {2023},
  editor = 	 {Krause, Andreas and Brunskill, Emma and Cho, Kyunghyun and Engelhardt, Barbara and Sabato, Sivan and Scarlett, Jonathan},
  volume = 	 {202},
  series = 	 {Proceedings of Machine Learning Research},
  month = 	 {7},
  publisher =    {PMLR},
  url = 	 {https://proceedings.mlr.press/v202/passaro23a.html},
}

@article{wang2024machine,
  title={Machine learning interatomic potential: Bridge the gap between small-scale models and realistic device-scale simulations},
  author={Wang, Guanjie and Wang, Changrui and Zhang, Xuanguang and Li, Zefeng and Zhou, Jian and Sun, Zhimei},
  journal={Iscience},
  volume={27},
  number={5},
  year={2024},
  publisher={Elsevier}
}

@article{batatia2023foundation,
      title={A foundation model for atomistic materials chemistry},
      author={Ilyes Batatia and Philipp Benner and Yuan Chiang and Alin M. Elena and Dávid P. Kovács and Janosh Riebesell and Xavier R. Advincula and Mark Asta and William J. Baldwin and Noam Bernstein and Arghya Bhowmik and Samuel M. Blau and Vlad Cărare and James P. Darby and Sandip De and Flaviano Della Pia and Volker L. Deringer and Rokas Elijošius and Zakariya El-Machachi and Edvin Fako and Andrea C. Ferrari and Annalena Genreith-Schriever and Janine George and Rhys E. A. Goodall and Clare P. Grey and Shuang Han and Will Handley and Hendrik H. Heenen and Kersti Hermansson and Christian Holm and Jad Jaafar and Stephan Hofmann and Konstantin S. Jakob and Hyunwook Jung and Venkat Kapil and Aaron D. Kaplan and Nima Karimitari and Namu Kroupa and Jolla Kullgren and Matthew C. Kuner and Domantas Kuryla and Guoda Liepuoniute and Johannes T. Margraf and Ioan-Bogdan Magdău and Angelos Michaelides and J. Harry Moore and Aakash A. Naik and Samuel P. Niblett and Sam Walton Norwood and Niamh O'Neill and Christoph Ortner and Kristin A. Persson and Karsten Reuter and Andrew S. Rosen and Lars L. Schaaf and Christoph Schran and Eric Sivonxay and Tamás K. Stenczel and Viktor Svahn and Christopher Sutton and Cas van der Oord and Eszter Varga-Umbrich and Tejs Vegge and Martin Vondrák and Yangshuai Wang and William C. Witt and Fabian Zills and Gábor Csányi},
      year={2023},
      eprint={2401.00096},
      journal={arXiv},
      primaryClass={physics.chem-ph}
}

@article{barroso_omat24,
  title   = {Open Materials 2024 (OMat24) Inorganic Materials Dataset and Models},
  author  = {Luis Barroso-Luque and Shuaibi Muhammed and Xiang Fu and Brandon M. Wood and Misko Dzamba and Meng Gao and Ammar Rizvi and C. Lawrence Zitnick and Zachary W. Ulissi},
  journal = {arXiv},
  year    = {2024}
}

@misc{levine2025openmolecules2025omol25,
      title={The Open Molecules 2025 (OMol25) Dataset, Evaluations, and Models}, 
      author={Daniel S. Levine and Muhammed Shuaibi and Evan Walter Clark Spotte-Smith and Michael G. Taylor and Muhammad R. Hasyim and Kyle Michel and Ilyes Batatia and Gábor Csányi and Misko Dzamba and Peter Eastman and Nathan C. Frey and Xiang Fu and Vahe Gharakhanyan and Aditi S. Krishnapriyan and Joshua A. Rackers and Sanjeev Raja and Ammar Rizvi and Andrew S. Rosen and Zachary Ulissi and Santiago Vargas and C. Lawrence Zitnick and Samuel M. Blau and Brandon M. Wood},
      year={2025},
      eprint={2505.08762},
      archivePrefix={arXiv},
      primaryClass={physics.chem-ph},
      url={https://arxiv.org/abs/2505.08762}, 
}

@misc{kaplan2025matpes,
  title         = {A Foundational Potential Energy Surface Dataset for Materials},
  author        = {Kaplan, Aaron D. and Liu, Runze and Qi, Ji and Ko, Tsz Wai and Deng, Bowen and Riebesell, Janosh and Ceder, Gerbrand and Persson, Kristin A. and Ong, Shyue Ping},
  year          = {2025},
  eprint        = {2503.04070},
  archivePrefix = {arXiv},
  doi           = {10.48550/arXiv.2503.04070}
}

@misc{PET-MAD-1.5-2026,
      title={High-quality, high-information datasets for universal atomistic machine learning},
      author={Cesare Malosso and Filippo Bigi and Paolo Pegolo and Joseph W. Abbott and Philip Loche and Mariana Rossi and Michele Ceriotti and Arslan Mazitov},
      year={2026},
      eprint={2603.02089},
      archivePrefix={arXiv},
      primaryClass={cond-mat.mtrl-sci},
      url={https://arxiv.org/abs/2603.02089},
}

@misc{UPET-2026,
      title={Pushing the limits of unconstrained machine-learned interatomic potentials},
      author={Filippo Bigi and Paolo Pegolo and Arslan Mazitov and Michele Ceriotti},
      year={2026},
      eprint={2601.16195},
      archivePrefix={arXiv},
      primaryClass={physics.chem-ph},
      url={https://arxiv.org/abs/2601.16195},
}

@inproceedings{
zhou2026matris,
title={Mat{RIS}: Toward Reliable and Efficient Pretrained Machine Learning Interatomic Potentials},
author={Yuanchang Zhou and Siyu Hu and Xiangyu Zhang and Hongyu Wang and Guangming Tan and Weile Jia},
booktitle={The Fourteenth International Conference on Learning Representations},
year={2026},
url={https://openreview.net/forum?id=5xBT5Ziute}
}

@inproceedings{fu2025learning,
  title={Learning Smooth and Expressive Interatomic Potentials for Physical Property Prediction},
  author={Fu, Xiang and Wood, Brandon M and Barroso-Luque, Luis and Levine, Daniel S and Gao, Meng and Dzamba, Misko and Zitnick, C Lawrence},
  booktitle={International Conference on Machine Learning},
  pages={17875--17893},
  year={2025},
  organization={PMLR}
}

\newpage
\appendix
\section{Reference Energy in DFT Dilemma}
\label{app:dft_notes}
In principle, density functional theory (DFT) provides a well-defined total energy for an electronic system within a given theoretical framework. For all-electron calculations, this total energy has an unambiguous physical meaning. In commonly used pseudopotential-based approaches, including the projector augmented-wave (PAW) method employed here, the absolute value of the total energy depends on the chosen atomic reference states and pseudopotentials, and is therefore not uniquely defined. As a result, the zero of energy becomes element-dependent and method-specific. Nevertheless, relative energies -- such as energy differences between configurations of the same system -- remain physically meaningful due to cancellation of these reference contributions.

The MACE architecture explicitly addresses this issue by decomposing the total energy into a sum of fixed isolated-atom reference energies and a learned interaction energy term. During training, only the interaction energy is modelled, while the isolated-atom energies for each element are kept fixed. This formulation enables training MACE models on data from multiple levels of theory simultaneously, a setting referred to as multi-head training. As a consequence, a specific head must be selected when fine-tuning a foundation model for a given task.

\begin{figure}[h]
\centering
\includegraphics[width=\textwidth]{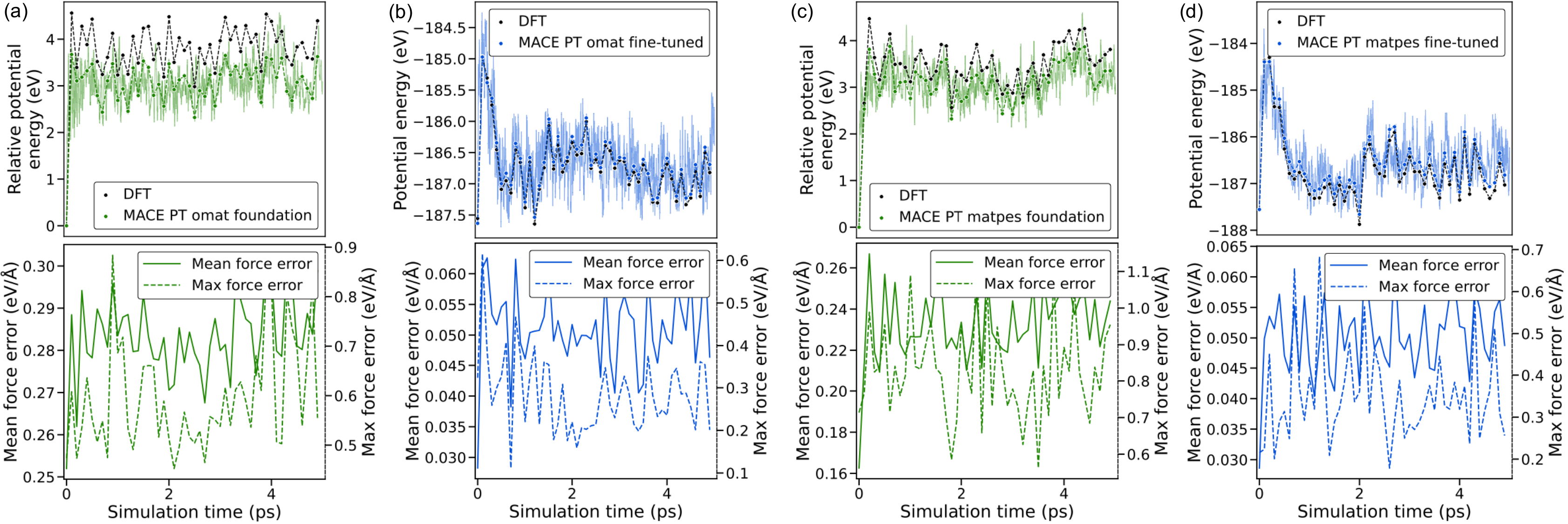}
\caption{Molecular dynamics simulations of CO$_2$ adsorbed on a fully OH-terminated MXene at 300 K, comparing different MACE foundation models. Panels (a,b) show MACE \textit{omat} foundation models (MACE PT \textit{omat}) without and with fine-tuning, respectively, while panels (c,d) show the corresponding models for the MATPES (MACE PT \textit{matpes}) foundation model. Each simulation consists of 10 000 time steps with a step size of 0.5 fs. At every 200th step, single-point DFT calculations are performed; top panels show DFT and ML potential energies along the trajectories, while bottom panels report the mean and maximum per-atom force L2 errors relative to DFT. For models without fine-tuning, ML potential energies are shown relative to the first configuration of each trajectory due to large, predominantly systematic energy offsets.}
\label{fig:MD CO2 comparison}
\end{figure}

\begin{figure}[h]
\centering
\includegraphics[width=\textwidth]{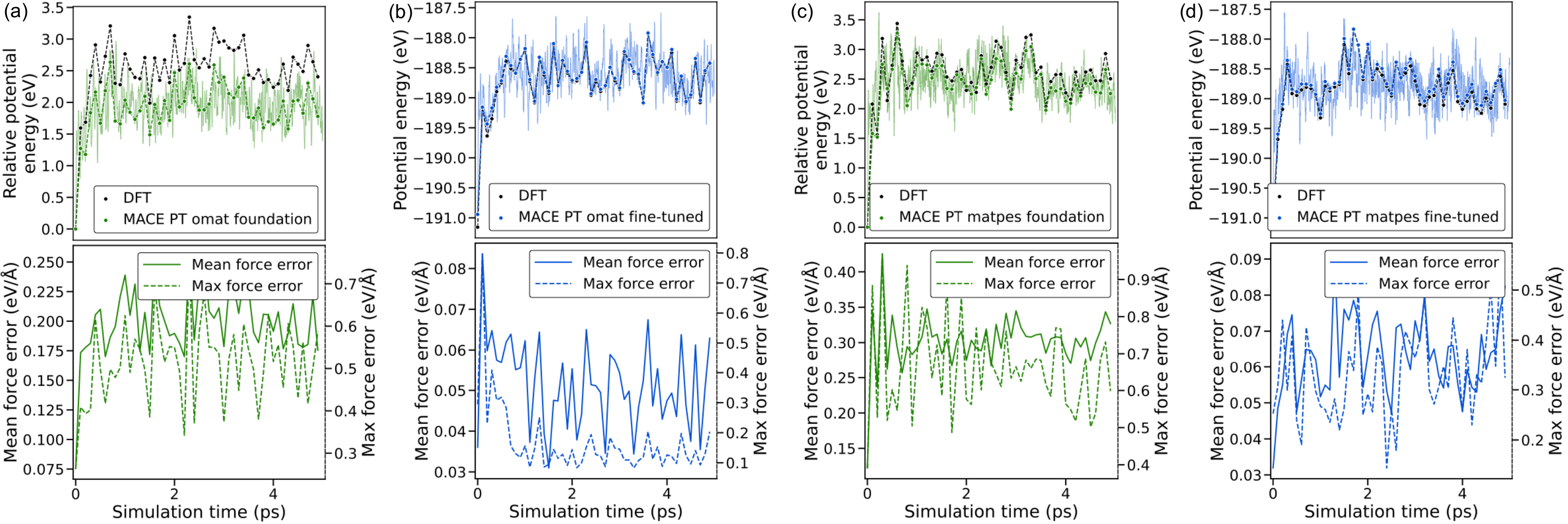}
\caption{Molecular dynamics simulations of HCOOH adsorbed on a fully O-terminated MXene at 300 K, comparing MACE models trained from different foundation models. Panels (a,b) show MACE \textit{omat} foundation models (MACE PT \textit{omat}) without and with fine-tuning, respectively, while panels (c,d) show the corresponding models for the MATPES (MACE PT \textit{matpes}) foundation model. Each simulation consists of 10 000 time steps with a step size of 0.5 fs. At every 200th step, single-point DFT calculations are performed; top panels show DFT and ML potential energies along the trajectories, while bottom panels report the mean and maximum per-atom force L2 errors relative to DFT. For models without fine-tuning, ML potential energies are shown relative to the first configuration of each trajectory due to large, predominantly systematic energy offsets.}
\label{fig:MD HCOOH comparison}
\end{figure}

In Figure.~\ref{fig:mace_intro}, we compare energies obtained from different foundation models after aligning them to our energy scale using differences in isolated-atom reference energies between our dataset and the respective foundation models. While this alignment significantly reduces the apparent energy offsets, substantial discrepancies remain, as evident in Figure~\ref{fig:mace_intro}. However, as shown in Figures~\ref{fig:MD CO2 comparison} and \ref{fig:MD HCOOH comparison}, relative energy variations within individual systems are less affected for the MACE PT \textit{matpes} and MACE PT \textit{omat} models. Even so, these differences are sufficiently large to motivate further training on system-specific data.

These observations also highlight the importance of the chosen level of theory, which in DFT is primarily determined by the exchange–correlation (XC) functional. 
A commonly used choice is the PBE functional, a generalised gradient approximation (GGA) in which the XC energy depends on the electron density and its gradient. In this work, we employ the rev-vdW-DF2 functional, which is also GGA-based but includes a non-local correlation term. This non-local contribution -- formally expressed as a double integral over the electron density and a kernel -- enables the description of van der Waals interactions, which are not captured by conventional GGAs by construction.

As a consequence, rev-vdW-DF2 introduces additional attractive interaction terms compared to standard GGA functionals such as PBE. 
Models trained on data generated with this functional therefore learn systematically more attractive interaction energies, which is consistent with the observed energy shifts relative to foundation models trained on datasets based on conventional GGAs.

While PBE and semi-empirical van der Waals corrections remain widely used, exploring alternative exchange–correlation functionals broadens the methodological landscape and allows assessment of model transferability across different physical descriptions. 
Rather than extending GGA through meta-GGA formulations, our approach adopts a complementary strategy by incorporating non-local correlation effects directly at the DFT level, and reflecting these consistently in the trained machine-learning interatomic potentials.
\begin{table*}[t]
\centering
\scriptsize
\setlength{\tabcolsep}{3.5pt}
\renewcommand{\arraystretch}{1.2}

\makebox[\textwidth][c]{%
\begin{tabular}{l c c c c c}
\toprule
\multirow[b]{2}{*}{\textbf{Model}} &
\multirow[b]{2}{*}{\makecell[c]{\textbf{Training}\\\textbf{time per run}\\\textbf{(GPU h)}}} &
\multicolumn{2}{c}{\textbf{Test MAE}} &
\multicolumn{2}{c}{\textbf{Test RMSE}} \\
\cmidrule(lr){3-4}\cmidrule(lr){5-6}
& &
\makecell{Energy/atom\\(meV) $\downarrow$} &
\makecell{Force\\(meV/\AA) $\downarrow$} &
\makecell{Energy/atom\\(meV) $\downarrow$} &
\makecell{Force\\(meV/\AA) $\downarrow$} \\
\midrule

EquiformerV2 \textit{Original} (30.8M) &
253.6 &
3.7 $\pm$ 0.2 (3.5) &
16.0 $\pm$ 0.2 (15.7) &
36.3 $\pm$ 5.7 (31.6) &
101.5 $\pm$ 3.6 (97.1) \\

EquiformerV2 \textit{Small} (4.8M) &
141.3 &
3.6 $\pm$ 0.1 (3.5) &
15.9 $\pm$ 0.7 (15.0) &
30.7 $\pm$ 1.3 (28.9) &
98.2 $\pm$ 3.1 (94.5) \\

\midrule

MACE (0.8M) &
234.3 &
2.1 $\pm$ 0.1 (2.0) &
15.5 $\pm$ 0.8 (14.8) &
19.5 $\pm$ 0.4 (19.0) & 
52.3 $\pm$ 15.6 (41.2)  \\

MACE \textit{PT omat} (6.2M) &
108.6* & 
1.8 $\pm$ 0.2 (1.7) &
14.6 $\pm$ 4.1 (11.7) &
19.5 $\pm$ 0.1 (19.4) & 
41.6 $\pm$ 5.9 (35.4)  \\

MACE\textit{ PT matpes} (0.8M) &
60.4* &
1.7 $\pm$ 0.1 (1.6) &
12.2 $\pm$ 0.2 (11.9) &
19.4 $\pm$ 0.5 (18.7) & 
36.0 $\pm$ 1.0 (35.1)  \\

\midrule
UPET cons. (2.9M) &
153.5 &
\makecell{ 1.3 $\pm$ 0.1 (1.2)} &
\makecell{20.6 $\pm$ 0.2 (20.4)} &
\makecell{18.6 $\pm$ 0.4 (18.2)} &
\makecell{50.3 $\pm$ 0.9 (49.5)} \\

UPET non-cons. (3.0M) &
72.1 &
\makecell{2.5 $\pm$ 0.1 (2.4)} &
\makecell{26.9 $\pm$ 1.6 (24.7)} &
\makecell{22.0 $\pm$ 1.0 (21.1)} &
\makecell{59.3 $\pm$ 3.2 (54.9)} \\

PET \textit{PT mad-s} (25.9M) &
36.5 &
\makecell{1.1 $\pm$ 0.0 (1.0)} &
\makecell{10.3 $\pm$ 0.0 (10.3)} &
\makecell{18.0 $\pm$ 0.0 (18.0)} &
\makecell{27.3 $\pm$ 1.3 (25.6)} \\

\midrule
MatRIS cons. (6.3M) &
 576.0 &
\makecell{ 4.8 $\pm$ 0.4 (4.4) } &
\makecell{ 23.5 $\pm$ 1.9 (21.6) } &
\makecell{ 19.8 $\pm$ 3.6 (17.1) } &
\makecell{ 170.3 $\pm$ 132.1 (74.1) } \\

MatRIS non-cons. (6.4M) &
 576.0 &
\makecell{ 5.8 $\pm$ 0.9 (4.8) } &
\makecell{ 22.2 $\pm$ 1.0 (21.0) } &
\makecell{ 32.2 $\pm$ 5.3 (26.6) } &
\makecell{ 130.3 $\pm$ 9.9 (119.2) } \\

MatRIS \textit{PT mp} (10.4M) &
 576.0 &
\makecell{ 7.0 $\pm$ 1.2 (5.9) } &
\makecell{ 35.7 $\pm$ 2.3 (33.7) } &
\makecell{ 45.8 $\pm$ 1.9 (43.6) } &
\makecell{ 381.5 $\pm$ 238.3 (198.2) } \\

MatRIS \textit{PT oam} (10.4M) &
 576.0 &
\makecell{ 5.2 $\pm$ 2.7 (3.4) } &
\makecell{ 18.4 $\pm$ 1.0 (17.6) } &
\makecell{ 18.7 $\pm$ 2.2 (17.1) } &
\makecell{ 81.3 $\pm$ 21.6 (56.7) } \\

\bottomrule
\end{tabular}
}

\caption{
Test performance of equivariant MLIPs. 
Reported are the MAE mean and standard deviation over three runs, with the best result in parentheses. Energy errors are reported per atom averaged over the total number of systems, and force errors as well. 
* indicates training on H100 instead of A100 (note that for EquiformerV2 and MACE, we used A100 FAT).
}
\label{tab:results}
\end{table*}

\section{Additional Quantitative Results}
\label{app:quantitative}

\cref{tab:results} summarises the test results of the benchmarked methods trained on the proposed dataset. 

On the original test set, EquiformerV2 \textit{Original} and EquiformerV2 \textit{Small} achieve very similar performance, with energy MAE of $3.7$ and $3.6$ meV/atom and force MAE of 16.0 and 15.9 meV/Å, respectively. 
The slightly lower standard deviation and RMSE of the \textit{Small} variant suggest improved robustness to outliers despite a substantially reduced parameter count, indicating that increasing EquiformerV2 capacity does not translate into improved accuracy for this dataset.

The 0.8M-parameter MACE model significantly improves energy accuracy relative to both EquiformerV2 variants, reducing energy MAE to $2.1$ meV/atom, while achieving comparable force MAE ($15.5$ meV/Å). 
Fine-tuning pretrained MACE foundation models leads to consistent gains. 
In particular, MACE \textit{PT matpes} achieves $1.7$ meV/atom energy MAE and $12.2$ meV/Å force MAE, outperforming the non-pretrained MACE at equal capacity. These results demonstrate that strong physical inductive biases combined with pretrained representations can compensate for limited model size and enable efficient transfer to MXene catalysis systems.

Among all evaluated models, PET \textit{PT mad-s} achieves the best overall performance on the original test set, with $1.1$ meV/atom energy MAE and $10.3$ meV/Å force MAE. 
Notably, these results are obtained with significantly lower training time than most other models, highlighting an advantageous accuracy–efficiency trade-off.

When evaluating generalisation to \fivetimesish larger systems (see~\cref{tab:cross_evaluation}), all models experience an increase in force MAE, while energy MAE increases more moderately or remains stable. EquiformerV2 variants show a pronounced degradation in force accuracy, reaching around $45–46$ meV/Å. 
The base MACE model and pretrained MACE variants exhibit improved robustness, with force MAE remaining in the $42–44$ meV/Å range, although pretrained variants do not uniformly improve force generalisation compared to the non-pretrained MACE. 
PET PT mad-s again performs best, achieving the lowest energy MAE ($1.3$ meV/atom) and force MAE ($36.3$ meV/Å) on the larger-system test set, indicating superior scalability and stability under increased system size.

Overall, these results show that 1) increased model capacity alone does not guarantee improved accuracy, 2) strong inductive biases and pretraining significantly benefit energy prediction accuracy, and 3) PET \textit{PT mad-s} provides the most consistent performance across both original and larger-system evaluations.

A direct comparison between conservative and non-conservative force-prediction variants highlights the impact of enforcing energy–force consistency. 
For UPET, the conservative model substantially improves force accuracy on the original test set, reducing force MAE from $26.9$ meV/Å (non-conservative) to $20.6$ meV/Å, while also achieving better energy MAE ($1.3$ vs. $2.5$ meV/atom). 
This indicates that constraining forces to be the gradient of the predicted energy provides a measurable benefit in accuracy on in-distribution configurations. 
On larger systems (see~\cref{tab:cross_evaluation}), both UPET variants experience a pronounced increase in force MAE, with the conservative model being slightly worse than the non-conservative one ($57.8$ vs. $49.4$ meV/Å), suggesting that conservative constraints alone do not guarantee improved extrapolation to larger system sizes.
For MatRIS, the conservative formulation outperforms the non-conservative variant on the original test set, achieving lower energy MAE ($4.1$ vs. $6.6$ meV/atom), with a slightly worse force MAE ($22.9$ vs. $22.7$ meV/Å). 
On larger systems, the conservative MatRIS model exhibits improved energy generalisation ($2.5$ vs. $3.9$ meV/atom), while force MAE remains comparable between the two variants. 
These results suggest that conservative force prediction systematically improves energy accuracy and physical consistency, while its impact on force MAE is model- and regime-dependent, with clear benefits observed on the original test distribution for some architectures.

Across both UPET and MatRIS, conservative models therefore show improved accuracy and physical consistency on in-distribution data, while gains in out-of-distribution generalisation depend on additional architectural and training factors beyond the conservative constraint alone.

\begin{figure}
\centering
\includegraphics[width=\textwidth]{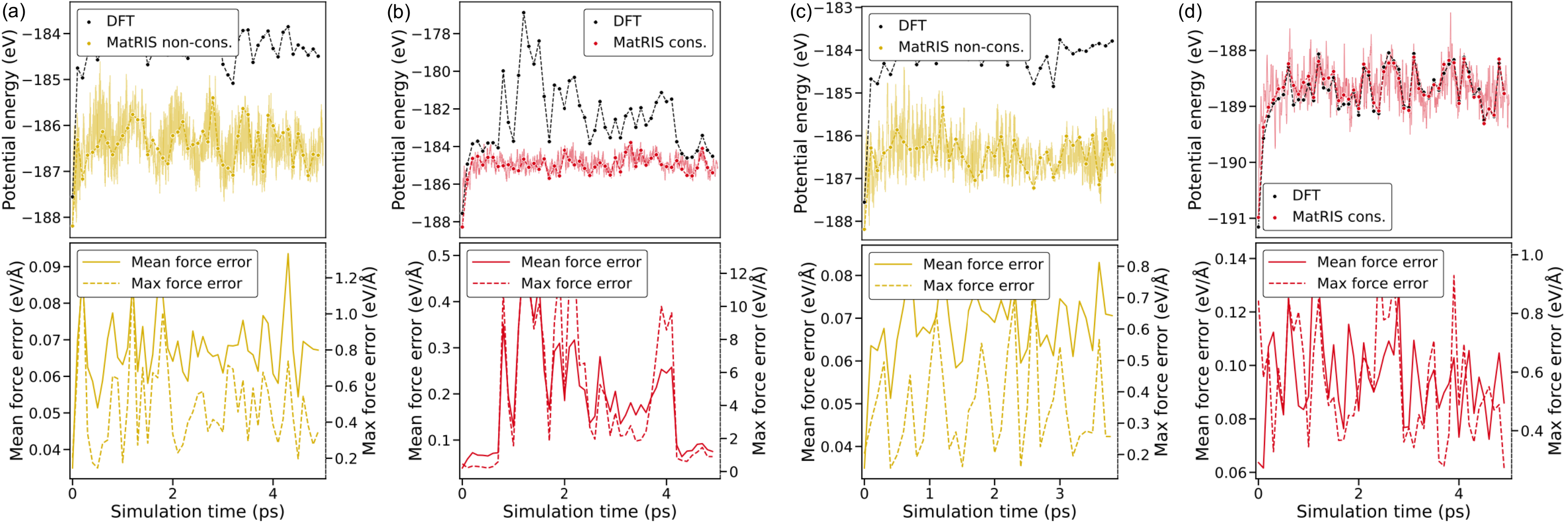}
\caption{Molecular dynamics simulations of (a,b) CO$_2$ on a fully OH-terminated MXene and (c,d) HCOOH on a fully O-terminated MXene at 300 K, comparing the performance of the (a,c) MatRIS non-cons. and (b,d) MatRIS cons. Each simulation consists of 10 000 time steps with a step size of 0.5 fs. At every 200th step, single-point DFT calculations are performed; top panels show the DFT and ML potential energies along the trajectories (defined with respect to elemental reference energies according to Eq.~\ref{eq:formation-energy}), while bottom panels report the mean and maximum per-atom force L2 errors relative to DFT.}\vspace{-10pt}
\label{fig:MD comparison_matris}
\end{figure}

\begin{figure}
\centering
\includegraphics[width=\textwidth]{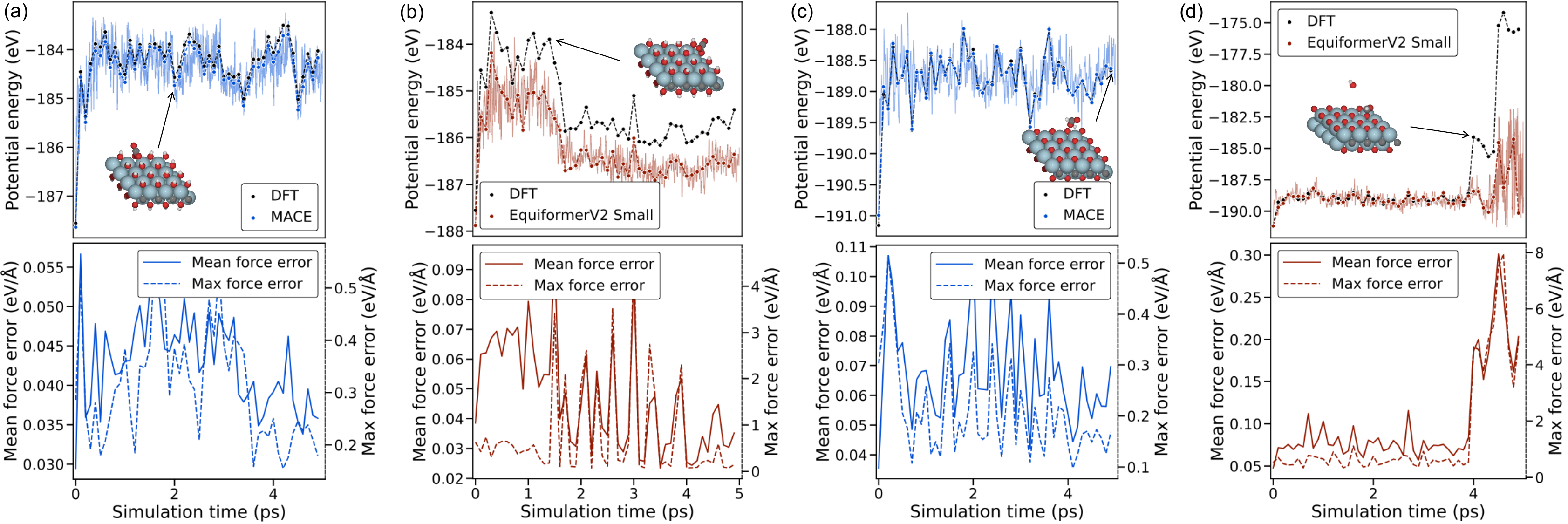}
\caption{Molecular dynamics simulations of (a,b) CO$_2$ on a fully OH-terminated MXene and (c,d) HCOOH on a fully O-terminated MXene at 300 K, comparing the performance of the (a,c) MACE and (b,d) EquiformerV2 models. Each simulation consists of 10 000 time steps with a step size of 0.5 fs. At every 200th step, single-point DFT calculations are performed; top panels show the DFT and ML potential energies along the trajectories (defined with respect to elemental reference energies according to Eq.~\ref{eq:formation-energy}), while bottom panels report the mean and maximum per-atom force L2 errors relative to DFT.}\vspace{-10pt}
\label{fig:MD comparison_mace-equi}
\end{figure}

\begin{figure}[h]
\centering
\includegraphics[width=0.75\textwidth]{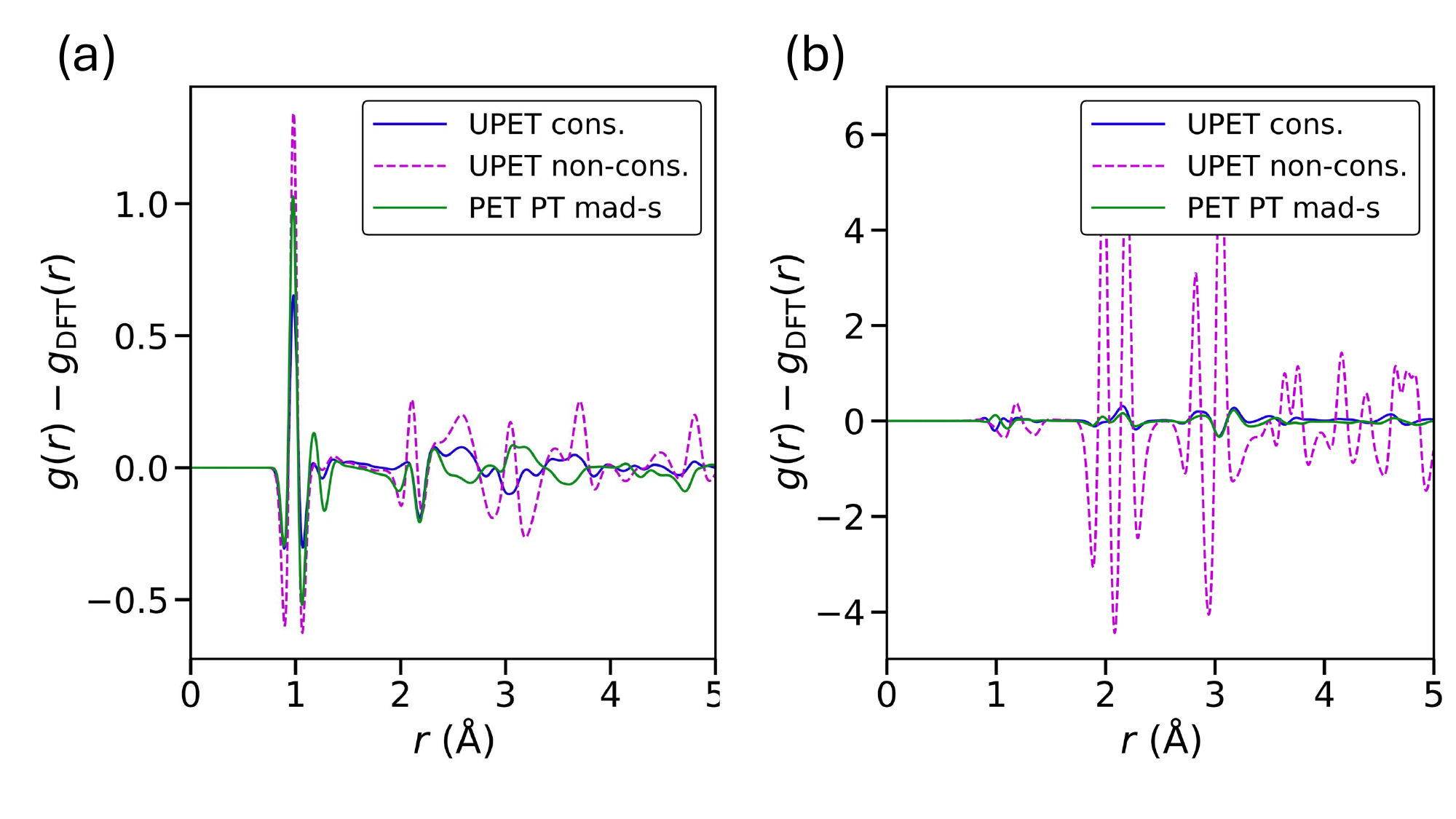}
\caption{Differences in radial distribution
functions between molecular dynamics 
simulations using machine learning (ML) 
models and density functional theory (DFT) 
for (a) CO$_2$ on a fully OH-terminated 
MXene and (b) HCOOH on a fully O-terminated
MXene. All simulations were carried out at 
300~K for 10\,000 steps with a time step of
0.5~fs.}
\label{fig:rdfs pets}
\end{figure}

\section{Additional Qualitative Results}
\label{app:qualitative}
Figures~\ref{fig:MD CO2 comparison} and \ref{fig:MD HCOOH comparison} show molecular dynamics (MD) simulations of
CO$_2$ and HCOOH, respectively, adsorbed on MXenes, evaluated using MACE foundation models both without (\ie zero-shot on MXene data) and with fine-tuning. 
To avoid large absolute energy offsets between the foundation models and the DFT reference -- even after correcting for differences in isolated-atom reference energies, as we accounted for in Figure~\ref{fig:mace_intro} -- the potential energies of the foundation models without any fine-tuning are shown relative to the initial configuration of each MD trajectory. 
This representation enables a meaningful assessment of relative energy variations along the trajectories.

Despite this favourable energy normalisation for the pre-trained foundation models, fine-tuning them using the proposed dataset leads to a clear improvement in the description of the energy evolution along the trajectories. Even more pronounced differences are observed for the forces, where the models without fine-tuning exhibit substantially larger errors. 
These results highlight the poor quality of existing foundation models on MXenes, and thus the need for the proposed dataset.

\end{document}